\documentclass{aa}  
\usepackage{graphicx}
\usepackage{amsmath}
\usepackage{amsfonts}
\usepackage{amssymb}
\usepackage{gensymb}
\usepackage{textcomp}
\usepackage[varg]{txfonts}
\usepackage[T1]{fontenc}
\usepackage{natbib}
\usepackage{xcolor}
\usepackage{hyperref}
\usepackage{placeins}

\bibpunct{(}{)}{;}{a}{}{,}
\makeatletter
\renewcommand*\aa@pageof{, page \thepage{} of \pageref*{LastPage}}
\makeatother

\begin{document}

   \title{Nature of small-scale Ca~{\sc ii}~H brightenings associated with weak fields in a quiet Sun region}

   \author{Xiang~Li \inst{1}
            \and
            H.~N.~Smitha \inst{1}
            \and
            S.~K.~Solanki \inst{1,2}
            \and
            T.~L.~Riethmüller \inst{1}
            \and
            L.~P.~Chitta \inst{1}
            \and
            S.~Jafarzadeh \inst{3,1}
          }

   \institute{Max-Planck-Institut für Sonnensystemforschung, Justus-von-Liebig-Weg 3, D-37077 Göttingen, Germany\\
              \email{lixiang@mps.mpg.de}
             \and
             School of Space Research, Kyung Hee University, Yongin, Gyeonggi 17104, Republic of Korea
             \and
             Astrophysics Research Centre, School of Mathematics and Physics, Queen's University Belfast, Belfast, BT7 1NN, UK
             }

   \date{Received xxx xxx, xxxx; accepted xxx xxx, xxxx}
  
  \abstract
   {Chromospheric emission observed in the Ca~{\sc ii}~H and K resonance lines is generally positively correlated with photospheric magnetic fields. However, a fraction of strong Ca~{\sc ii}~H\&K brightenings are found to occur over regions of low photospheric magnetic flux density.}
   {We aim to understand the origins of these chromospheric brightenings that are associated with weak fields in quiet Sun regions.}
   {We use high resolution and high cadence observations in the Ca~{\sc ii}~H line taken by the Sunrise Filter Imager (SuFI) along with the photospheric magnetograms recorded by the Imaging Magnetograph eXperiment (IMaX), instruments onboard the first flight of the balloon-borne observatory {\scshape Sunrise}. We also make use of the atmospheric physical parameters inverted from the IMaX data by the Stokes-Profiles-INversion-O-Routines (SPINOR) code.}
   {217 Ca~{\sc ii}~H brightenings associated with weak fields are identified. 111 of these are located at the peripheries of strong field regions, in the magnetic canopies formed by the expansion of the kG flux tube. The other brightenings are associated with either weak unipolar magnetic patches (25 brightenings), with interactions between weak mixed-polarity magnetic fields (75 brightenings), or at locations where no clear magnetic features can be identified (6 brightenings). For the latter two situations, the Ca~{\sc ii}~H intensity is found to vary with a period of 3--5 minutes.  This is accompanied by variations in the line-of-sight velocity, the temperature at $\log\tau = -2.5$, and the magnetic flux with a similar period.}
   {About half of the strong Ca~{\sc ii}~H brightenings associated with weak fields are located in magnetic canopies overlying very weak photospheric fields. These canopies are produced by the expansion of kG flux tubes. About 10-12\%\ are associated with small magnetic bright points where the magnetic field is probably not spatially resolved, so that it appears weak, while likely reaching kG strength in reality. The remainder appears to be Ca~{\sc ii}~H grains, which, however, are overwhelmingly found to be associated with magnetic flux emergence or cancellation events.  We conclude that although Ca~{\sc ii} grains are mainly driven by (magneto-)acoustic shock waves, the observations indicate that interactions between magnetic fields with opposite polarities also contribute to the formation of the brightest peaks.}

   
   \titlerunning{Nature of small-scale Ca~{\sc ii}~H brightenings associated with weak fields in quiet Sun region}
   \authorrunning{X.~Li et al.}
   \maketitle
   \nolinenumbers
%
\section{Introduction}
\label{section:Intro}

The solar chromosphere is highly dynamic, with complex structures revealed by observations at various wavelengths from the ultraviolet to millimeter wavebands \citep{Solanki2004}. In a standard plane-parallel model of the solar atmosphere, the temperature starts to increase through the chromosphere above the temperature minimum region \citep[e.g.,][]{Vernazza1981}. This enhancement in temperature and the underlying heating mechanism operating in the chromosphere are still being debated \citep{Carlsson2019}.

One widely recognized heating mechanism in the quiet chromosphere is acoustic shock heating, which is inferred from short-lived brightenings in the Ca~{\sc ii}~K/H wavebands. These brightenings are usually referred to as Ca~{\sc ii}~K/H grains \citep{Beckersphd}. Observations reveal that their sizes vary from sub-arcsec to a couple of arcsec, and that they are periodic with typical periods of 2--4 min (e.g., \citealp{Cram1977,Rutten1991,Kneer1993,Beck2008,Mathur2022}). Various studies have tried to reproduce this phenomenon through simulations. For instance, using one-dimensional radiation-hydrodynamic numerical simulations under non–local thermodynamic equilibrium (NLTE) conditions, \citet{Carlsson1992,Carlsson1997} simulated the Ca~{\sc ii}~K line profile with vertically propagating sinusoidal acoustic waves. They found that waves with a periodicity of 3\,minutes can reproduce the Ca~{\sc ii}~K line, and its temporal evolution quite well, and that the bright grains are formed by shocks located around 1\,Mm above the surface (corresponding to an optical depth unity at 5000$\,\AA$, or $\tau_{5000}=1$). \citet{Wedemeyer2004} ran a three-dimensional radiation hydrodynamic simulation, finding that some of the upward-propagating acoustic waves produce shocks in the chromosphere also in 3D due to a drop in the gas density.

The magnetic nature of the chromospheric grains is still under debate. On the one hand,  \citet{Lites1999} did not find any clear evidence for the presence of  magnetic features at the locations of the chromospheric grains, and thus concluded that there is no direct correlation between them. This is also supported by other studies (e.g., \citealp{Remling1996,Worden1999,Beck2008}), who decided that the Ca~{\sc ii}~K/H grains are non-magnetic in nature. On the other hand, \citet{Sivaraman1982} found a one-to-one correlation between the Ca~{\sc ii}~$K_{2V}$ bright points and their photospheric magnetic counterparts. Later, \citet{Sivaraman2000} also found that some chromospheric bright points are associated with merging and cancellation of magnetic features. Recently, \citet{MartinezGonzalez2023} reported Ca~{\sc ii}~K grains occurring when the associated magnetic flux reaches its maximum, and that these grains reappear in around 4 minutes, indicating that they might be manifestations of magnetohydrodynamic waves traveling upward through the atmosphere. 

In general, most previous research has been based on observations which either have low spatial resolution, and/or last only a short period of time. To better understand the relationship between the photospheric magnetic field and the chromospheric grains, observations with high spatial resolutions as well as longer time series are necessary, as pointed out by \citet{Kamio2006,MartinezGonzalez2023}.

While it is still unclear whether the magnetic field plays a role in the formation of chromospheric grains, the photospheric magnetic fields are considered to be crucial for the structuring of the chromosphere. This is reflected in the strong positive correlation between the chromospheric emission (commonly seen in the~Ca {\scshape ii}~K/H line at 393.3/396.8 nm) and the photospheric magnetic flux density in both quiet-Sun (QS) and active regions (ARs), which indicates that significant chromospheric brightenings are more likely to appear in strong-field regions. Quantitative analysis of this relationship has been carried out by a number of authors over the past decades, who found a power-law or logarithmic function to well describe it (e.g., \citealp{Schrijver1989,Harvey1999,Ortiz2005,Loukitcheva2009,Kahil2017}). Although most chromospheric brightenings are located in strong-field regions, some of them show strong deviations from this relationship, i.e., they are found to occur over relatively weak fields in the photosphere \citep{Schrijver1989}. To our knowledge, no detailed study has so far investigated their origins. Consequently, the nature of chromospheric heating mechanisms responsible for these brightenings in the weak field regions remains unclear. We speculate that many of the Ca~{\sc ii}~H brightenings studied here are related to Ca~{\sc ii} grains, so that our results may give insight into the nature of these features.

In this paper, we present an observational study using the high spatial resolution data acquired during the first flight of the balloon-borne stratospheric solar observatory {\sc Sunrise} \citep{Solanki2010,Barthol2011,Berkefeld2011}. We investigate the nature of the Ca~{\sc ii}~H brightenings associated with weak fields in a QS region and interpret the results in terms of Ca~{\sc ii}~H grains. 
In Section \ref{section:Data}, we describe the data used in this study, as well as the detailed data reduction steps. In Section \ref{section:identification}, we introduce the criterion for the Ca~{\sc ii}~H brightenings associated with weak fields. In Section \ref{section:result}, we present our results, showing the characteristics of the Ca~{\sc ii}~H brightenings we identified. In Section \ref{section:discussion}, we discuss our results, putting forward possible explanations for these brightenings. Our conclusions are presented in Section \ref{section:conclusion}. 

\section{Observations and data preparations}
\label{section:Data}

We use the observations acquired by the {\sc Sunrise} observatory during its first science flight starting on 2009 June 9. {\sc Sunrise} is composed of a 1.0 m aperture Gregory telescope, equipped with two post-focus instruments: the Sunrise Filter Imager \citep[SuFI; ][]{Gandorfer2011} and the Imaging Magnetograph eXperiment \citep[IMaX; ][]{Martinez2011}. The observational target  was a quiet-Sun region close to disk center ($\mu=0.97$). These datasets consist of three time series recorded between 00:36 -- 00:58 UT, 01:30 -- 01:59 UT, and 14:22 -- 14:43 UT on 2009 June 9.

   \subsection{SuFI and IMaX data}
   \label{subsection:SuFI and IMaX data}
   
   The ultraviolet (UV) imager SuFI has five wavelength channels: 214\,nm\ (bandwidth of 10\,nm), 300\,nm\ (bandwidth of 5\,nm), 312\,nm\ (OH-band, bandwidth of 1.2\,nm), 388\,nm\ (CN-band, bandwidth of 0.8\,nm), and 397\,nm\ (core of the Ca~{\sc ii}~H line, bandwidth of 0.18\,nm). In this paper, we only used the SuFI observations at 300\,nm\ and 397\,nm, which sample the lower photosphere and lower chromosphere, respectively. The cadence is 12 s (for time series 00:36 -- 00:58 UT and 01:30 -- 01:59 UT) and 39 s (for time series 14:22 -- 14:43 UT). The field-of-view (FOV) of SuFI is 15$\times$40 arcsec$^2$, and the corresponding image scale is 0.02\arcsec\,pixel$^{-1}$, which varies slightly in different wavebands. We employed the level 3 data, which were reconstructed by the phase diversity technique using an averaged wavefront obtained from the permanent in-flight recordings of a focused and a defocused image \citep{Hirzberger2010,Hirzberger2011}.

   Simultaneously with SuFI, the IMaX acquired spectropolarimetric data by scanning the photospheric Fe\,{\sc i}  spectral line at $\lambda_0$ = 5250.2\,\AA, which is highly sensitive to magnetic field with a Land\'e factor $g = 3$. It recorded full Stokes maps at five wavelength positions ($-80, -40, +40, +80, +227$ m$\AA$ from the line center), with a spectral resolution of 85 m$\AA$. The FOV of IMaX is 50$\times$50 arcsec$^2$, and the image scale is 0.055\arcsec\,pixel$^{-1}$. We used the phase-diversity reconstructed data (level 2), with a noise level of $3\times10^{-3}I_{\rm c}$ (where $I_{\rm c}$ is the continuum intensity). This corresponds to $\sigma_{B}\approx$ 14\,G in the Line-of-sight (LOS) photospheric magnetic field \citep{Martinez2011}. The observing mode is V5-6 (full Stokes vector recorded in 5 wavelengths and with 6 images accumulated per wavelength point), with a cadence of 33\,s.

   \begin{figure*}
     \centering
     \includegraphics[width=\hsize]{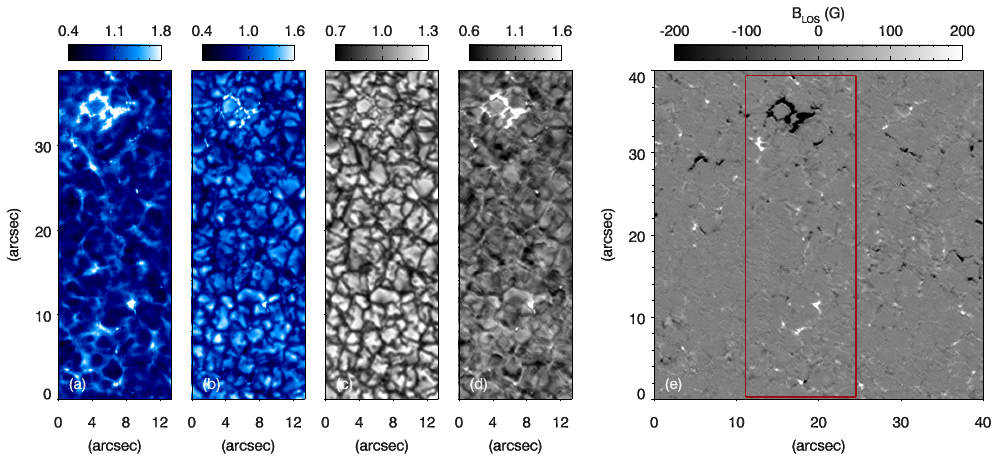} 
     \caption{Spatially aligned SuFI and IMaX images, and photospheric magnetogram obtained from IMaX data. The images correspond to (a) SuFI Ca~{\sc ii}~H line core at 397 nm, (b) SuFI at 300 nm, (c) IMaX continuum at 5250.4 $\AA$, (d) IMaX line core at 5250.2 $\AA$, and (e) whole FOV of the magnetogram. $Panel \ a - d$ are contrast images. See the main text for a detailed description of the intensity contrast. The maroon box in $panel \ e$ encloses the common FOV of SuFI and IMaX.}
     \label{fig:oridata} 
   \end{figure*}

   \subsection{Inversions}
   \label{subsection:inversions}

   To better understand the nature of small-scale Ca\,{\sc ii}\,H brightenings, we consider the photospheric parameters, temperature, magnetic field vector, and LOS velocity. For this, we use the inversion code SPINOR (Stokes-Profiles-INversion-O-Routines \citep{Frutigerphd,Frutiger2000}), which is based on STOPRO routines to solve the Unno-Rachkovsky equations under LTE conditions \citep{Solankiphd}. A simple one-component atmospheric model with three optical depth nodes (at log $\tau = -2.5, -0.9, 0$, where $\tau$ is the optical depth at a wavelength of 5000\,\AA) for the temperature, and one node each for  magnetic field strength, inclination, azimuth, LOS velocity and micro-turbulence was used as the initial guess atmospheric model. The code first computes synthetic spectra of  Fe\,{\sc i} 5250.2\AA\ from the initial atmospheric model. Then the atmosphere is tweaked by changing the free parameters until a good fit to the observed  Stokes profile in the spatial pixel in question is achieved. The atmosphere which produces the best fit to the full Stokes profiles is taken to be the representative model giving rise to the observed spectra. More details on the inversion strategy are described by \citet{Solanki2017}. The velocity map was corrected for the etalon blueshift caused by the collimated setup \citep{Martinez2011}. Note that the output $B_{\rm LOS}$ and $v_{\rm LOS}$ correspond roughly to the values at $\log\tau = -1$ \citep{Riethmuller2017}.

   \subsection{Image alignment}
   \label{section:alignment}

   Alignment among datasets from SuFI and IMaX is essential before doing a detailed analysis. We followed the steps below to align the images:
   \begin{enumerate}
      \item Cut off the edges of IMaX images, which are fuzzy due to the apodization. The usable FOV of IMaX images was reduced to 40$\times$40 arcsec$^2$.
      \item Remove image jitter in the SuFI 397\,nm, SuFI 300\,nm, and IMaX Stokes $I$ continuum time series using a cross-correlation technique. Then apply the resulting IMaX offsets to other wavelength positions and Stokes parameters of IMaX as well as the SPINOR inversion maps.
      \item Compare the SuFI 397 nm, SuFI 300 nm, and IMaX Stokes $I$ continuum by eye for a preliminary alignment. At first, randomly pick images at one moment in time, and clip them to a roughly common FOV. Then visually align these images according to some common features (the bright points in SuFI 300 nm and 397 nm; the granulation in SuFI 300 nm and IMaX Stokes $I$ continuum).
      \item Use a cross-correlation technique to achieve a more accurate alignment. Resample the SuFI images (plate scale $0.02''$ pixel$^{-1}$) to the same plate scale as IMaX (plate scale $0.055''$ pixel$^{-1}$). Then align the SuFI 300 nm bandpass with IMaX Stokes $I$ continuum by applying offsets acquired from a cross-correlation technique, and use the same procedure to align the SuFI 397 nm bandpass with IMaX Stokes $I$ line core ($\lambda_0 = 5250.2 \,\AA$). Here the IMaX Stokes $I$ line core maps are obtained by picking the minimum intensity near the Fe~{\sc i} line center in the inverted Stokes $I$ profile from the SPINOR inversion. Figure \ref{fig:oridata}d shows an example of the obtained line core map (clipped to the SuFI FOV).
   \end{enumerate}

   Figure \ref{fig:oridata}a--d show an example of SuFI contrast images at 397 nm, 300 nm, as well as IMaX Stokes $I$ continuum and line core images after being aligned, with a common FOV of $14''\times39''$. The corresponding magnetogram is also displayed in Figure \ref{fig:oridata}e. In all the four intensity images, the intensity is normalized to the mean QS intensity $I_{\rm qs}$ at the corresponding wavelength. The normalized intensity is called contrast ($C_{\rm WB}$), following \citet{Kahil2017}:

   \begin{equation}
     \label{eq:contrast}
     C_{\rm WB} = \frac{I_{\rm WB}}{I_{\rm WB, qs}},
   \end{equation}
   where WB = \{CON, LC, 300, 397\} represents four wavelength bands: ``CON'' and ``LC'' respectively stand for the IMaX Stokes $I$ continuum and line core, and ``300'' and ``397'' are SuFI 300 nm and 397 nm wavelength bands respectively. $I_{\rm WB, qs}$ is the mean QS intensity averaged over the whole FOV (for IMaX data it is $40''\times40''$, for SuFI data it is $14''\times39''$).

\section{Identification of Ca~{\sc ii}~H brightenings associated with weak fields}
\label{section:identification}

\begin{figure}
     \centering
     \includegraphics[width=\hsize]{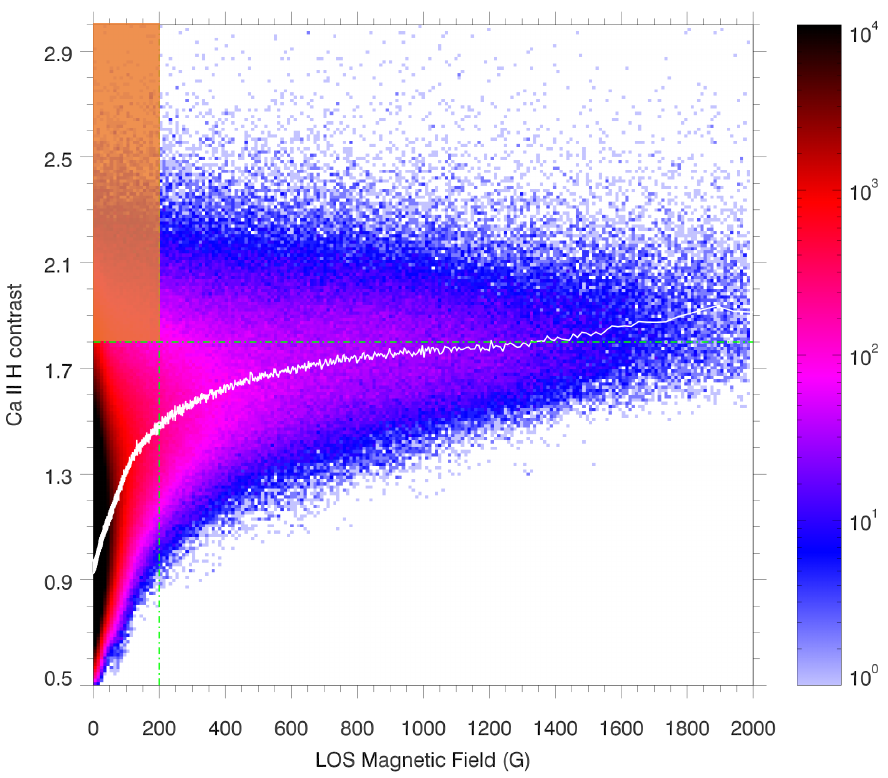} 
     \caption{Scatterplot of the Ca~{\sc ii}~H contrast vs. the photospheric LOS magnetic field. The white curve is the binned contrast curve with each bin containing 500 data points. The orange rectangular area corresponds to the excess Ca~{\sc ii}~H brightenings associated with weak fields. The dashed horizontal green line marks the lower intensity contrast limit (1.8) of the brightenings, and the dashed vertical green line marks the upper $B_{LOS}$ limit (200 G) of the brightenings considered in this paper.}
     \label{fig:criterion} 
   \end{figure}

   To identify the brightenings associated with weak fields, we made use of the scatterplot of the contrast in the SuFI 397\,nm waveband versus the photospheric $B_{LOS}$ shown in Figure~\ref{fig:criterion}. The binned contrasts were calculated as the average value in each bin containing 500 data points. They are overplotted as a white curve, which highlights the positive correlation between the Ca~{\sc ii}~H intensity and the photospheric magnetic field. In regions with $|B_{\rm LOS}|<200\,{\rm G}$, there is a large scatter in chromospheric contrast and in particular, the data points sampled by the orange rectangular area strongly deviate from the binned contrast curve. We chose these as the Ca~{\sc ii}~H brightenings associated with weak fields. The criterion for selecting these brightenings is:
   \begin{equation}
     \label{eq:criterion1}
     C_{397} > 1.8 \ {\rm with} \ |B_{\rm LOS}| < 200 \ \mathrm{G}, 
   \end{equation}
   where the value 1.8 is set as the lower limit of the intensity contrast of Ca~{\sc ii}~H brightenings, since it is roughly $4\sigma_{397}$ above the mean Ca~{\sc ii}~H intensity over the whole FOV of SuFI (the average Ca~{\sc ii}~H contrast is equal to 1; $\sigma_{397}\approx0.2$ is the standard deviation of Ca~{\sc ii}~H contrast, and this value varies very little across different images). We set 200 G as the upper/lower limit of weak/strong $B_{\rm LOS}$, since this value is well below the equipartition field strength in the photosphere ($\sim300-500$\,G; e.g., \citealp{Lin1995,Khomenko2003,Orozco2007,Martinez2008}). We note that weak field here refers to pixel-level $|B_\mathrm{LOS}|<200\,\mathrm{G}$, not necessarily weak intrinsic field strength. In the text that follows, all references to Ca~{\sc ii}~H brightenings, if not otherwise stated, are referred to as brightenings identified by the above criterion. 

\begin{figure}
     \centering
     \includegraphics[width=\hsize]{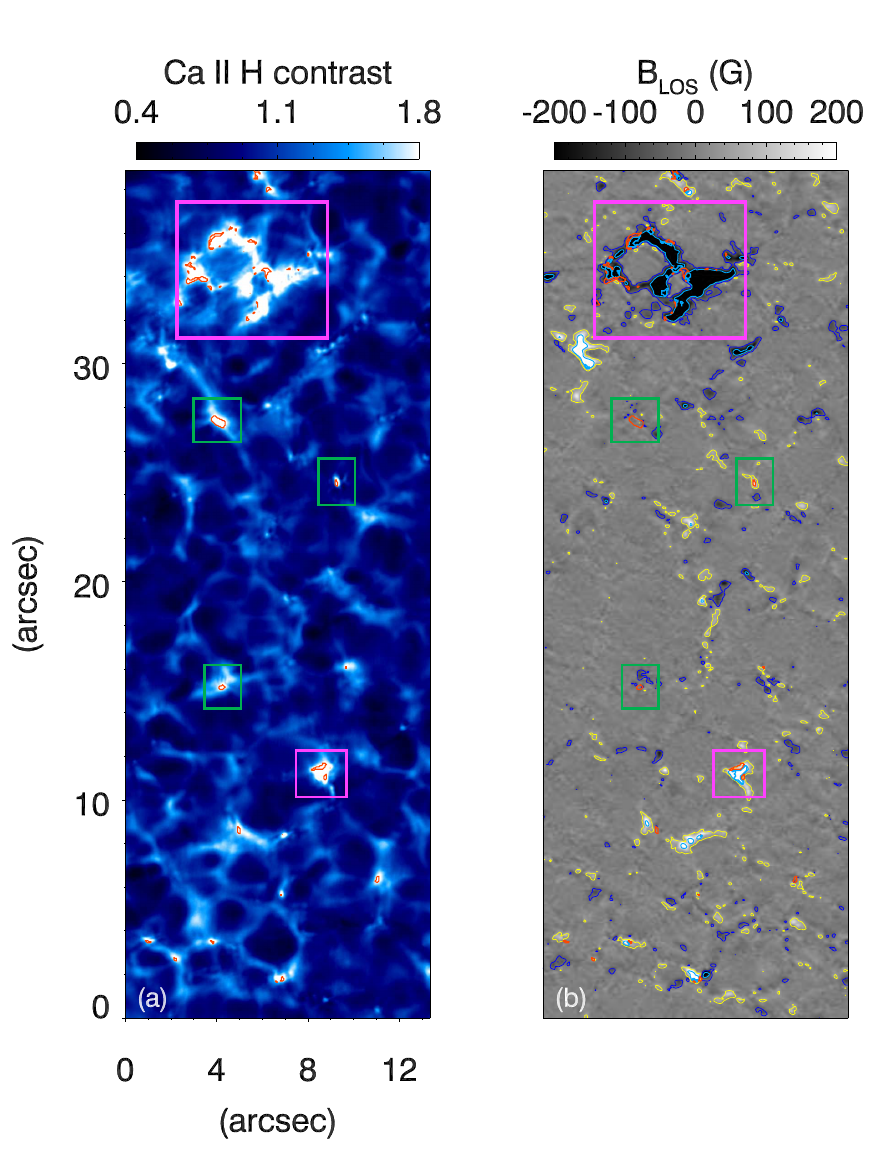} 
     \caption{An example of  brightenings identified to be lying the green rectangle in Fig.~\ref{fig:criterion} overlaid on the Ca~{\sc ii}~H image ($panel \ a$) and co-aligned IMaX LOS magnetogram ($panel \ b$). The Ca {\sc ii} brightenings associated with weak fields are marked with orange contours. The yellow and blue contours in the magnetogram respectively enclose positive- and negative-polarity magnetic features with $|B_{LOS}|$ above 3$\sigma_{B}$ (40 G). The cyan contours enclose strong field regions with $|B_{LOS}| > 200 \ \mathrm{G}$. The pink and green rectangles highlight brightenings associated with strong and weak-field features. See the main text for more details.}
     \label{fig:overview_brightening} 
   \end{figure}

   Figure~\ref{fig:overview_brightening} shows a Ca~{\sc ii}~H image and the corresponding co-aligned photospheric magnetogram. We used a threshold of 40 G, approximately 3 times of the noise ($3\sigma_B$), as the lower limit for identifying a magnetic feature. Only few Ca~{\sc ii}~H brightenings associated with weak fields are identified, and most of these brightenings are restricted to small spatial scales. This is consistent with the fact that very few data points in the scatterplot satisfy the brightening criterion. Figure~\ref{fig:overview_brightening} also reveals some characteristics of the locations of these small-scale brightenings:

   \begin{itemize}
       \item some of them are located in weak field regions (e.g., the brightenings in the green rectangles);
       \item some of them are located at the peripheries of strong field regions (e.g., the brightenings in the pink rectangular regions).
   \end{itemize}

   These characteristics indicate that the Ca~{\sc ii}~H brightenings associated with weak fields are related to different types of magnetic features, which possibly implies a difference in their natures. To get a clearer picture of the physical processes and properties behind these brightenings, it is necessary to analyze their evolution in combination with other physical parameters obtained from IMaX and SuFI data. When selecting and categorizing brightenings for detailed analysis, some criteria were set as follows.

   \begin{itemize}
   \item[(1)]Only the brightenings lasting for at least two frames are considered. In general, if two individually identified brightenings in two consecutive frames share at least one spatial pixel, then they are taken as the same brightening observed at two points in time.
   \item[(2)]The brightenings that touch the spatial boundaries of the image at any time during their evolution are removed. This is because in this case, we miss the part of the brightening that is outside the FOV, so we cannot get an accurate picture of their nature.
   \item[(3)]The brightenings that still exist at either the first or last frame of a time sequence are also removed, since we do cannot follow the entire evolution of these brightenings and their corresponding magnetic features, which could mislead our understanding of them.
   \item[(4)]If a brightening splits into several brightenings, or a number of  brightenings merge into one single brightening, then we consider them as a single brightening.
   \item[(5)]All the brightenings that touch the boundary of a strong field region (with $|B_{LOS}| > 200 \,\mathrm{G}$) at any time during their lifetime are considered to be brightenings in close proximity to strong field regions. Otherwise the brightening is considered to be located in a weak field region.
   \end{itemize}

   Since the total length of the time series is relatively short (only 134 frames in total) and the total number of identified Ca~{\sc ii}~H brightenings is small, we did the inspection manually. In total, 217 Ca~{\sc ii}~H brightenings associated with weak fields were selected. Among them, 106 brightenings are located in weak field regions ($|B_{LOS}|$ < 200 G also in the vicinity of the studied brightenings), while the other 111 brightenings are in close proximity to strong field regions ($|B_{LOS}|$ > 200 G).

\section{Results}
\label{section:result}

   \subsection{Brightenings located in weak magnetic field regions}
   \label{subsection:brightenings_weak_field}

   The brightenings located in weak magnetic field regions are associated with different types of magnetic features. We classified these brightenings into three classes:

   \begin{itemize}
   \item Class 1: Brightenings in unipolar magnetic patches.
   \item Class 2: Brightenings associated with the interaction between two or more mixed-polarity magnetic patches.
   \item Class 3: Brightenings in areas where no magnetic patches are identified.
   \end{itemize}

   \begin{table}[h]
     \centering
     \begin{tabular}{cc}
       \hline\hline
       Category & Number of cases (percentage)\\
       \hline
       Class 1 & 25 (23.6\%)\\
       Class 2 & 75 (70.8\%)\\
       Class 3 & 6 (5.7\%)\\
       \hline
     \end{tabular}
     \caption{Results from the identification of Ca~{\sc ii}~H brightenings associated with weak fields.}
     \label{tab:class_stat}
   \end{table}

   Among all the 106 brightenings located exclusively in weak field regions, the numbers and fractions of different categories are given in Table~\ref{tab:class_stat}. Clearly, 
   there are only very few cases where no clear magnetic patches are identified (i.e., Class 3), which means that most of the brightenings are associated with diverse magnetic features resolved by {\sc Sunrise} data. The overwhelming fraction of them belong to Class 2, which illustrates that the brightenings mainly occur when magnetic patches interact.

   In the next subsections, we will describe some typical examples of each class in detail.

   \begin{figure*}[h!]
     \centering
     \includegraphics[width=0.9\hsize]{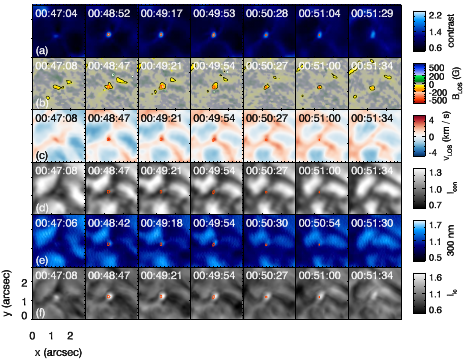}
     \caption{Time series of a typical Class 1 Ca~{\sc ii}~H brightening. $From$ $top$ $to$ $bottom$ $row$: Ca~{\sc ii}~H intensity contrast at 397 nm ($panel \ a$); line-of-sight magnetic field, $B_{LOS}$ ($panel \ b$); line-of-sight velocity, $v_{LOS}$ ($panel \ c$); normalized continuum intensity at 5250.4 \AA\ ($panel \ d$); SuFI 300 nm contrast ($panel \ e$); Fe~{\sc i} 5250.2 \AA\ line core intensity contrast ($panel \ f$). The Ca~{\sc ii}~H brightening, as well as magnetic patches with positive and negative polarities are marked in the same way as in Figure~\ref{fig:case_strongfield}.}
     \label{fig:case_class1}
   \end{figure*}

   \subsubsection{Class 1}
   \label{subsubsection:class1}

   Figure~\ref{fig:case_class1} displays the evolution of a Ca~{\sc ii}~H brightening occurring in a unipolar magnetic patch. The brightening, appearing as a very small point-like feature, is identified from 00:48:52 to 00:51:04 UT. This brightening is located in a small negative-polarity magnetic structure with $|B_{LOS}| < 200 \,\mathrm{G}$, and associated with a strong downflow. In the intensity maps of the Fe~{\sc i} continuum, 300 nm waveband, and Fe~{\sc i} line core (Fig.~\ref{fig:case_class1}d, e, f), the bright point is also visible, although only rather faintly in the continuum. The barycenter of this brightening shifts slightly at different heights, which may be due to the inclination of the magnetic element \citep{Jafarzadeh2014}. These features indicate that the bright point is a MBP.

   The actual field strength of a MBP is on the order of kilogauss (kG) \citep{Nagata2008, Lagg2010, Riethmuller2014}. It means that in this case, the $B_{LOS}$ of the magnetic patch is probably underestimated, which makes the Ca~{\sc ii}~H brightening in Figure~\ref{fig:case_class1} appear as a brightening associated with a weak field. Considering that the size of the bright point in Ca~{\sc ii}~H waveband is very small ($\sim0.2''$), the underestimation of the field strength is likely because the true diameter of the kG magnetic element in the lower photosphere is below the spatial resolution of {\sc Sunrise}/IMaX ($0.15''$--$0.18''$) \citep{Jafarzadeh2013,Riethmuller2014}.

   \subsubsection{Class 2}
   \label{subsubsection:class2}

   Most of the Ca~{\sc ii}~H brightenings associated with weak fields are accompanied by multiple magnetic patches. These features can be either two distinct opposite-polarity patches, or a variety of tiny, mixed-polarity patches.

   \paragraph{Brightenings associated with magnetic features of two distinct magnetic patches}
   \label{paragraph:class2_two_mag}
   
   \
   
\begin{figure*}[h!]
     \centering
     \includegraphics[width=\hsize]{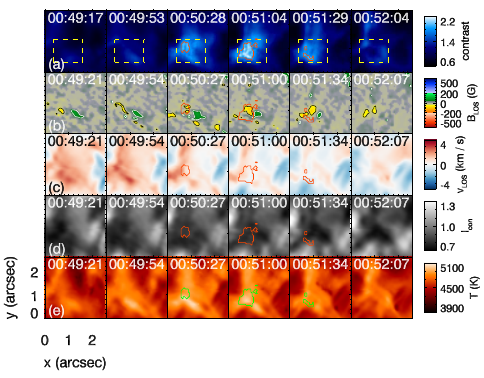}
     \caption{Evolution of a Class 2 Ca~{\sc ii}~H brightening that is associated with a magnetic cancellation event. $From \ top \ to \ bottom \ row$: Ca~{\sc ii}~H contrast ($panel \ a$); photospheric $B_{LOS}$ map ($panel \ b$); photospheric $v_{LOS}$ ($panel \ c$); normalized continuum intensity at 5250.4 $\AA$ ($panel \ d$); temperature at $\log\tau = -2.5$ ($panel \ e$). Magnetic patches are marked as in Figure~\ref{fig:case_class1}. The Ca~{\sc ii}~H brightening is enclosed by orange contours in $panel \ a-d$, and green contours in $panel \ e$. The dashed yellow box in $panel \ a$ highlights the region of the Ca~{\sc ii}~H brightening, which will be used for quantitative analysis of other physical parameters in Figure~\ref{fig:case_class22_variation}.}
     \label{fig:case_class2_2}
   \end{figure*}

   Changes in magnetic flux associated with the Ca~{\sc ii}~H brightenings are dominated by flux cancellation or emergence between two distinct magnetic patches. Figure~\ref{fig:case_class2_2} shows the evolution of a brightening over its lifetime, associated with a magnetic cancellation event. In the first two frames (00:49:17 -- 00:49:53 UT), no obvious brightening can be seen in the Ca~{\sc ii}~H bandpass, while the magnetograms show that there are two magnetic features (with the maximum $B_{LOS}\approx 100 \,\mathrm{G}$) moving toward each other. Then, from 00:50:28 UT, a bright patch becomes visible in Ca~{\sc ii}~H, until it disappears at 00:52:04 UT. In the meantime, a cancellation event happens below the brightening: from 00:50:27 UT, the flux of the negative polarity feature constantly decreases, until it disappears in the last frame (00:52:07 UT). The flux of the positive polarity feature also slightly decreases overall, although there are multiple splittings and mergings in this process.

   As chromospheric brightening and magnetic field cancellation proceed, the $v_{LOS}$ in this region switches from strong downflow to a weak upflow. The granules, especially those on the lower-left, slightly move together, which might drive the motion of the magnetic features. Throughout the evolution of this Ca~{\sc ii}~H brightening, no obvious response is found in continuum maps. Nevertheless, the temperature at $\log\tau = -2.5$ shows a similar variation as that in Ca~{\sc ii}~H intensity maps: it first rises, then drops back.

   \begin{figure}
     \centering
     \includegraphics[width=\hsize]{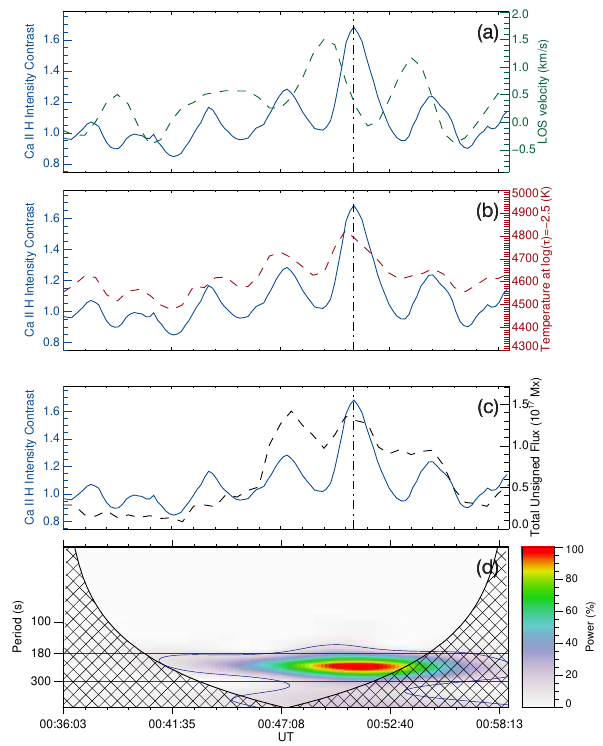}
     \caption{Comparisons between temporal variation of Ca~{\sc ii}~H intensity contrast and that of other physical parameters in the region enclosed by the yellow dashed box in Figure~\ref{fig:case_class2_2}. The temporal variation of Ca~{\sc ii}~H contrast is plotted as a blue solid curve in $panel \ a - c$. The overplotted curves are the averaged line-of-sight velocity (green dashed curve in $panel \ a$), average temperature at $\log\tau = -2.5$ (red dashed curves in $panel \ b$), and total unsigned magnetic flux (black dashed curve in $panel \ c$). The vertical black dot-dashed line in $panel \ a - c$ marks the time 00:51:04 UT, when the Ca~{\sc ii}~H intensity reaches its maximum. $Panel \ d$ shows the wavelet power spectra of the Ca~{\sc ii}~H intensity variation, where the power is normalized to its maximum and shown in percentages. The cross-hatched area is the cone-of-influence (COI). The black contour marks the 95\% confidence level. The dark green horizontal solid lines indicate periods of 3 min and 5 min, respectively.}
     \label{fig:case_class22_variation}
   \end{figure}

   In this case, the strong Ca~{\sc ii}~H brightening can be identified only from 00:50:28 to 00:51:29 UT.. However, after examining a longer time series, we found that the transient excess Ca~{\sc ii}~H brightness occurs periodically at the same location, and there are also some correlations with other physical parameters. In Figure~\ref{fig:case_class22_variation}, we compare the temporal variation (00:36:03 -- 00:58:41 UT) of averaged Ca~{\sc ii}~H intensity contrast with that of averaged $v_{LOS}$, averaged temperatures at $\log\tau = -2.5$, and total unsigned magnetic flux inside the yellow dashed box region in Figure~\ref{fig:case_class2_2}a. When calculating the magnetic flux, we included all the pixels with $|B_{LOS}|$ stronger than 14 G ($1\sigma_B$) instead of 40 G ($3\sigma_B$), to obtain a more accurate magnetic flux in this region.

   As shown in Figure~\ref{fig:case_class22_variation}, there is a clear periodic oscillation in the Ca~{\sc ii}~H intensity contrast. We computed the Morlet wavelet power spectrum of the Ca~{\sc ii}~H time series using WaLSAtools implementation \citep{Jafarzadeh2025} in Fig~\ref{fig:case_class22_variation}d, where the signal was linearly detrended and apodized with a Tukey window ($\alpha$ = 0.1) prior to the wave analysis. From the wavelet analysis, we obtained a dominant period of 3-4 min, which is consistent with the dominant chromospheric oscillation period \citep{Rutten1991}. This periodic variation can also be found in the line-of-sight velocity in Fig~\ref{fig:case_class22_variation}a, which precedes the Ca~{\sc ii}~H contrast by $\sim 1$ min. In Fig~\ref{fig:case_class22_variation}b, we can see that the temperature at $\log\tau = -2.5$ shows a clear correlation with the change in the Ca~{\sc ii}~H intensity, and a slight time shift between them can be seen. The strong correlation indicates that the Ca~{\sc ii}~H brightening reflects the temperature enhancement in the lower chromosphere, and the time shift is consistent with the time it takes an upward propagating acoustic or magneto-acoustic wave to travel between $\log\tau = -2.5$ and the formation height of Ca~{\sc ii}~H. 

   The variations in Ca~{\sc ii}~H intensity and the magnetic flux are compared in Fig~\ref{fig:case_class22_variation}c. During the occurrence of the brightening, a slightly decreasing trend is found in the total unsigned magnetic flux, while it is still around a local maximum. For the previous and the subsequent period, the magnetic flux also reaches a local maximum together with the Ca~{\sc ii}~H intensity at 00:48 UT and 00:55 UT, respectively. In general, the variation in magnetic flux indicates that magnetic activity plays a role in the occurrence of the excess Ca~{\sc ii}~H intensity. On the other hand, it is also possible that the magnetic activity and the intensity variations are modulated by 3-min waves propagating from below into the chromosphere.

   One typical example of a Ca~{\sc ii}~H brightening associated with a flux emergence event is shown in Appendix~\ref{appendix:A}, where the variation of the Ca~{\sc ii}~H intensity, $v_{LOS}$, temperatures at $\log\tau=-2.5$, and magnetic flux are similar to the event displayed above.

   \paragraph{Brightenings associated with a number of tiny, mixed-polarity patches}
   \label{paragraph:class2_mix_mag}

   \

   \begin{figure*}[h!]
     \centering
     \includegraphics[width=0.9\hsize]{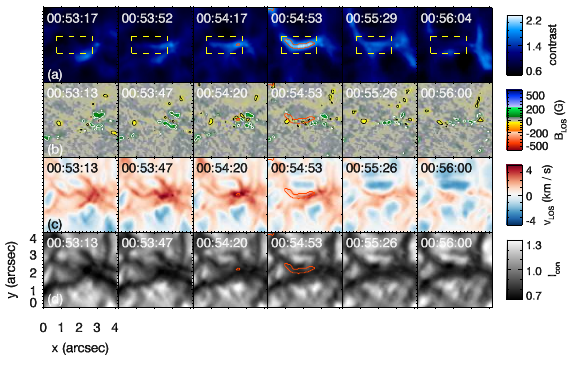}
     \caption{Evolution of a Class 2 Ca~{\sc ii}~H brightening associated with a number of tiny, mixed-polarity magnetic patches. Ca~{\sc ii}~H line core intensity maps ($panel \ a$), magnetograms ($panel \ b$), $v_{LOS}$ maps ($panel \ c$), and 5250.4 $\AA$ continuum maps ($panel \ d$) are displayed. The yellow dashed box encloses the area over which the curves plotted in Fig.~\ref{fig:case_class23_variation} are averaged.}
     \label{fig:case_class2_3}
   \end{figure*}

   The occurrence of some Ca~{\sc ii}~H brightenings is accompanied by the interactions between multiple tiny, mixed-polarity magnetic features. Figure~\ref{fig:case_class2_3} presents one example of this kind of brightening. The brightening, highlighted inside the dashed yellow box in $panel \ a$, has a slender shape, occurring in an intergranular lane. As the Ca~{\sc ii}~H intensity increases from 00:53:52 -- 00:54:53 UT (identified as a brightening from 00:54:17 to 00:54:53 UT), more tiny magnetic patches with mixed polarities appear, distributed exactly where the brightening is located. These magnetic patches are rather scattered and show little persistence over time. Then after 00:55:29 UT, the Ca~{\sc ii}~H intensity decreases and the brightening is no longer identified. The $v_{LOS}$ maps show a downflow that gradually gets stronger (00:53:17 -- 00:54:17 UT) before the brightening occurs. Later, the downflow becomes weaker, with a few surrounding upflow regions intruding into the downflow region. While the brightening is visible in the Ca~{\sc ii}~H bandpass, no obvious brightening can be seen in the continuum at the photosphere.

   In this case, the region hosting the brightening is bustling with short-lived small-scale weak magnetic patches of mixed polarities, with $|B_{LOS}|>40\,\mathrm{G}$. The appearance and disappearance of magnetic features happens in each frame, which indicates that flux emergence or concentration and cancellation may happen nearly simultaneously in this region, so that either of these processes may contribute to the Ca~{\sc ii}~H brightening. However, these magnetic processes are not sufficiently clear (e.g., magnetic features just appear in one frame and then disappear in the next frame), probably because these magnetic patches are so small that the flux emergence/concentration and cancellation only last for a rather short period of time that is below the cadence of our data (33 s).

   \begin{figure}[h!]
     \centering
     \includegraphics[width=1\hsize]{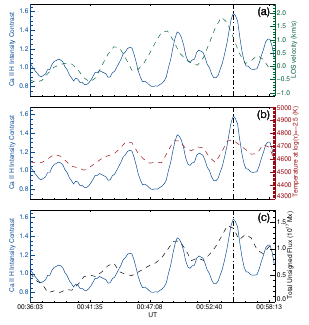}
     \caption{Same as Figure~\ref{fig:case_class22_variation} but for the Ca~{\sc ii}~H brightening case shown in Figure~\ref{fig:case_class2_3}. The vertical black dot-dashed line in each panel marks the time moment 00:54:53 UT, when the Ca~{\sc ii}~H intensity reaches the maximum.}
     \label{fig:case_class23_variation}
   \end{figure}
   Figure~\ref{fig:case_class23_variation} displays the temporal variation of corresponding physical quantities averaged over the brightening region from 00:36:03 UT to 00:58:41 UT. The Ca~{\sc ii}~H brightening in this case is also part of the 3--5 min intensity oscillation, which is further strengthened by the wavelet analysis in App.~\ref{appendix:B}. The peak value of the Ca~{\sc ii}~H intensity shows an increasing trend until it reaches its maximum at 00:54:53 UT (the vertical dot-dashed line). This periodic variation can also be found in the photospheric $v_{LOS}$ and temperature at $\log\tau = -2.5$, which correlates well with the change in the Ca~{\sc ii}~H intensity.

   The variation in the magnetic flux also shows a time dependence similar to the Ca~{\sc ii}~H intensity (see Figure~\ref{fig:case_class23_variation}c). We first focus on the time period around 00:54:53 UT. Before the Ca~{\sc ii}~H intensity reaches the peak, the total unsigned magnetic flux increases by $\sim7\times10^{16}$ Mx starting from 00:52 UT. The increasing magnetic flux could reflect the newly emerging flux from the interior, or external flux brought into the box region, or partly concentrated by the strong downflows to rise above the noise and become visible in IMaX data. Then, the magnetic flux decreases along with a drop in the Ca~{\sc ii}~H intensity. Interestingly, this positive correlation between the variations in the Ca~{\sc ii}~H intensity and the magnetic flux is also present over the whole time series. Thus, from 00:41:35 UT to 00:54:53 UT, the Ca~{\sc ii}~H intensity and the magnetic flux, both show an overall increasing trend. 

   \subsubsection{Class 3}
   \label{subsubsection:class3}

   For a few Ca~{\sc ii}~H brightenings, we could not find any associated magnetic patches with signals above the noise (i.e. fulfilling our criterion of 40 G). These brightenings are grouped into Class 3.

   \begin{figure*}[h!]
     \centering
     \includegraphics[width=0.9\hsize]{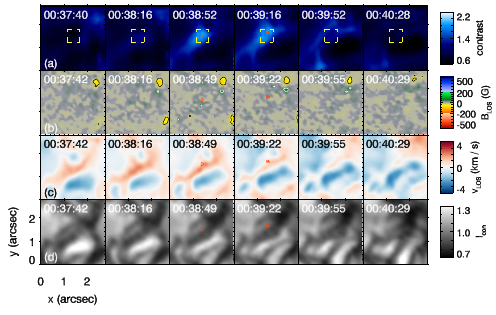}
     \caption{Same as Figure~\ref{fig:case_class2_3} but for the evolution of a Class 3 Ca~{\sc ii}~H brightening.}
     \label{fig:case_class3}
   \end{figure*}

   \begin{figure}[h!]
     \centering
     \includegraphics[width=1\hsize]{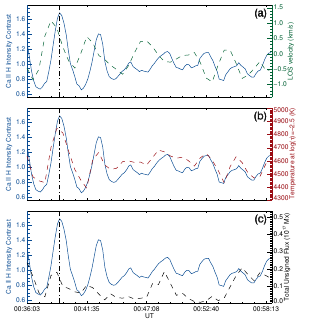}
     \caption{Same as Figure~\ref{fig:case_class22_variation} but for the Ca~{\sc ii}~H brightening shown in Figure~\ref{fig:case_class3}. The vertical black dot-dashed line in each panel marks the time 00:38:52 UT, when the Ca~{\sc ii}~H intensity reaches its peak.}
     \label{fig:case_class3_variation}
   \end{figure}

    One example of Class 3 Ca~{\sc ii}~H brightenings is given in Figure~\ref{fig:case_class3}. The brightening appears instantly in the intergranular lane at 00:38:52 UT, existing for only two frames. In the magnetogram, no magnetic patches are identified close to the brightening. The $v_{LOS}$ maps show a clear downflow before the brightening occurs, and after that, the downflow weakens, with two upflow regions nearby growing larger and moving close to each other.
   
   Figure~\ref{fig:case_class3_variation} presents the temporal evolution of corresponding averaged physical parameters of this brightening case. Ca~{\sc ii}~H intensity, line-of-sight velocity, and temperature at $\log\tau = -2.5$ all show periodic variations as in the previous cases. The total unsigned magnetic flux rises to a local maximum of $1.5\times10^{16}$ Mx at 00:38 UT, right before the Ca~{\sc ii}~H intensity reaches a maximum. Then the flux decreases, and increases to another local maximum at 00:49 UT, which corresponds to another Ca~{\sc ii}~H intensity peak. The oscillation of the magnetic flux in this case is not very clear, which might be because that the magnetic field is too weak to reveal clear magnetic features above the noise.
   
   In this case, the field strength is quite weak. The lack of small-scale magnetic features means that, either they play no role in producing the Ca~{\sc ii}~H brightening, or are not fully resolved by our magnetic field observations, or their evolution time is shorter than 33 s, the cadence of the magnetograms. Given that the median lifetime of the internetwork magnetic features is 1.1 min \citep{Anusha2017}, it is more likely that such events are not affected by the magnetic field, although we cannot completely rule out that photospheric magnetic fields still play a role in causing these brightenings.

   \subsection{Brightenings located at the periphery of strong magnetic field regions}
   
    \begin{figure*}[h!]
     \centering
     \includegraphics[width=0.9\hsize]{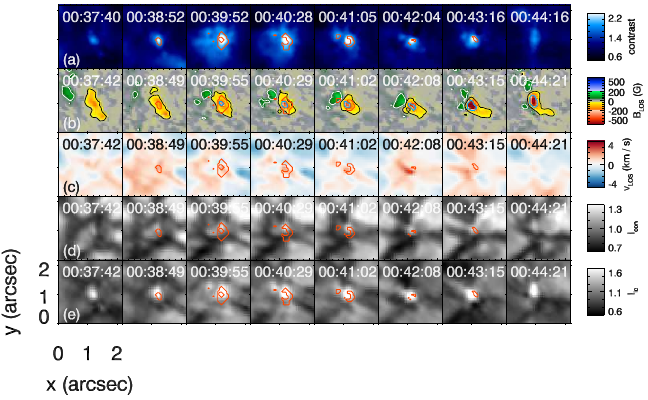}
     \caption{Temporal sequence of an identified Ca~{\sc ii}~H brightening located at the periphery of a strong field region. $From \ top \ to \ bottom \ row$: Ca~{\sc ii}~H 397 nm intensity recorded by {\sc Sunrise}/SuFI ($panel \ a$); line-of-sight (LOS) component of photospheric magnetic field, $B_{LOS}$ ($panel \ b$); photospheric line-of-sight velocity, $v_{LOS}$ ($panel \ c$); normalized continuum (+227 m\AA\ from the Fe~{\sc i} 5250.2 \AA\ line core) intensity ($panel \ d$); Fe~{\sc i} 5250.2 \AA\ line core intensity contrast ($panel \ e$). The orange contour encloses the identified Ca~{\sc ii}~H brightening. The white/black contours in $panel \ b$ mark positive/negative magnetic polarity patches. The cyan contour in $panel \ b$ encloses a strong field region with $|B_{LOS}| > 200 \,\mathrm{G}$.}
     \label{fig:case_strongfield}
   \end{figure*}

   \begin{figure}[h!]
     \centering
     \includegraphics[width=\hsize]{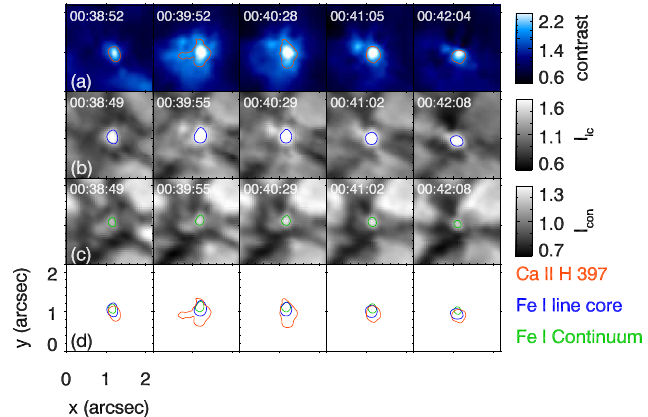}
     \caption{Comparison of the Magnetic Bright Point (MBP) in Figure~\ref{fig:case_strongfield} at wavelengths corresponding to different heights. The contours of the bright point in the Ca~{\sc ii}~H line core ($panel \ a$), Fe~{\sc i} line core ($panel \ b$), and continuum ($panel \ c$) intensity images are overplotted respectively in orange, blue, and green. These contours are drawn overlapping in $panel \ d$ to compare their sizes and locations.}
     \label{fig:case_bp_size}
   \end{figure}
   \label{subsection:brightenings_strong_field}
   
   Some of the identified Ca~{\sc ii}~H brightenings associated with weak fields actually occur in the immediate vicinity of strong fields. Figure~\ref{fig:case_strongfield} shows one example where an identified brightening associated with weak fields is located at the periphery of a strong magnetic field region. In this example, we only focus on the brightening identified above the negative-polarity magnetic patch. The region with an excessive Ca~{\sc ii}~H intensity has a circular shape, located in a strong negative-polarity magnetic patch. In addition to the Ca~{\sc ii}~H images, the circular brightening is also visible in Fe~{\sc i} line core and continuum images. These facts indicate that this is a magnetic bright point (MBP) \citep{Nagata2008,Riethmuller2010}. 

   The brightening becomes apparent at 00:38:52 UT. It appears inside a negative-polarity magnetic patch. Then at 00:39:52 UT, the area with a strong Ca~{\sc ii}~H intensity grows, while the $B_{\rm LOS}$ at the center of the magnetic patch increases to more than 200\,G, and the contour of the brightening associated with weak fields is concentric around the central strong field region ($|B_{LOS}|$ > 200 G). From this point on, the identified brightening is exactly located at the periphery of the negative-polarity magnetic patch, where the magnetic field is relatively weak ($|B_{LOS}|$ < 200 G). This topology is maintained for more than 3 minutes until 00:44:16 UT, when the Ca~{\sc ii}~H intensity decreases and no peripheral brightening is identified anymore, although a central brightening overlying the magnetic feature remains. In the $v_{LOS}$ maps, a downflow can be clearly seen at the location of the Ca~{\sc ii}~H brightening all the time.

   In this example, the strong Ca~{\sc ii}~H intensity originates from the strong negative-polarity magnetic patch, but extends beyond the strong-field feature, overlying weak fields. It is worth noting that the contour of the identified lower chromospheric brightening lies like a partial ring around the MBP for most of the time. This is consistent with the expansion of a strong-field flux tube with height.

   The expansion of the flux tube can be verified if the size of the MBP is found to be larger at greater heights \citep{Jafarzadeh2014}. We plotted the contours of the MBP in Ca~{\sc ii}~H, Fe~{\sc i} line core, and continuum images in Figure~\ref{fig:case_bp_size}. These three wavebands respectively sample the lower chromosphere, the middle-to-upper photosphere, and the lower photosphere. For determining the contour of the MBP, we adopted a criterion similar to the one used in \citet{Riethmuller2008}: in each waveband and each frame, we enclosed all pixels with an intensity contrast $C$ meeting the following criterion
   \begin{equation}
       \label{eq:criterionMBP}
       C > \frac{C_{max}+1}{2},
   \end{equation}
   where $C_{max}$ is the maximum intensity contrast of the MBP, and the value unity is the background averaged intensity contrast. The value $\frac{C_{max}+1}{2}$ roughly corresponds to the $FWHM$ of the intensity profile of the bright point.

   We can see that the size of the MBP increases from the Fe~{\sc i} continuum to the line core to the Ca~{\sc ii}~H bandpass, especially in Fig.~\ref{fig:case_bp_size}d, where the contours of the MBP are plotted on top of each other. Note that the SuFI Ca~{\sc ii}~H filter integrates over a wide height range, and the obtained Ca~{\sc ii}~H intensity has a large contribution from the photosphere \citet{Jafarzadeh2013}, so the difference of the MBP in size between the Ca~{\sc ii}~H and Fe~{\sc i} line core observations is not significant at some times. However, the size of the MBP in the continuum is significantly smaller than that observed in the other two wavebands all the time. This result supports previous conclusions that the cross-section of the magnetic flux tube expands with height \citep[e.g.,][]{Spruit1976,GrossmannDoerth1988,Solanki1989,Solanki1990,Buehler2015}. Since magnetic field lines at greater heights overlie quiet regions on the surface \citep{Jafarzadeh2017b}, when viewing the MBP vertically from above, the outer part of the MBP in the lower chromosphere is projected beyond the MBP in the photosphere (i.e., outside the strong field region), so that the outer part of the Ca~{\sc ii}~H bright point is associated with weak fields. Furthermore, as seen in Fig.~\ref{fig:case_bp_size}d, the MBP is slightly shifted to the lower right at higher layers, suggesting that the flux tube is inclined in this direction \citep[see also][]{Jafarzadeh2014}. This is the likely cause for only the right half of the MBP in Ca~{\sc ii}~H images being identified as a brightening associated with weak fields. Accordingly, this kind of brightenings is probably still physically associated with the strong-field magnetic element nearby.

\section{Discussions}
\label{section:discussion}

From the above results we can see that in the QS, the Ca~{\sc ii}~H brightenings associated with weak fields are related to different types of magnetic features. We note that, as defined in Sect.~\ref{section:identification}, we used the co-aligned photospheric magnetograms to classify the underlying magnetic environment of the Ca~{\sc ii}~H brightening, and track their temporal evolutions. We do not claim that there is a precise pixel-by-pixel correspondence between the Ca~{\sc ii}~H brightening and the intrinsic magnetic field. Our main findings are summarized below. 

A large fraction of the seemingly weak-field Ca~{\sc ii}~H brightenings are located at the peripheries of strong field regions ($|B_{LOS}|$ > 200 G). They are actually the outer part of the bright areas that originate from strong fields in the photosphere, while directly overlying pixels having only weak magnetic fluxes due to the expansion of (likely kG) magnetic flux tubes, i.e., the magnetic canopy. Since the Fe~{\sc i} line recorded by IMaX samples both the magnetic atmosphere above the base of the canopy as well as the field-free or relatively weakly magnetized atmosphere below the canopy, it often effectively senses only a relatively weak field. We propose that these brightenings are physically associated with strong fields.

Another group of brightenings, the Class 1 events, are seemingly associated with weak, $|B_{LOS}|$ < 200 G, fields. However, these fields produce brightenings in all the layers observed by {\sc Sunrise} during its 2009 flight, looking like classical magnetic bright points. We interpret them as kG magnetic elements that are not fully resolved even at the spatial resolution achieved by IMaX on {\sc Sunrise} \citep[cf.][]{Riethmuller2014}.

The most common class of brightenings located in what are likely truly weak field regions (with $|B_{LOS}|$ at or below equipartition values), the Class 2 events, are associated with the interaction between mixed-polarity magnetic patches. At their sites magnetic cancellation/disappearance and emergence/appearance events are found. Finally, a small fraction of the brightenings are classified as Class 3 events, which do not appear to be related to concrete magnetic structure, or these fields are very weak ($|B_{LOS}| < 40 \ \mathrm{G}$) or spatially unresolved.

The rapidly varying field, composed of mainly small, weak, short-lived magnetic features, found to be associated with the Ca~{\sc ii}~H brightenings in Class 2 events may be the product of a small-scale turbulent dynamo acting in the near-surface layers of the Sun \citep{Vogler2007, Rempel2014, Bhatia2022}, evidence for which has been presented by various authors \citep{Danilovic2010, Buehler2013, Lites2014}.

The Ca~{\sc ii}~H brightenings grouped in Class 2 and 3 are transient. We checked the variations of the Ca~{\sc ii}~H intensity and the photospheric LOS velocity in the corresponding region, and found that all of them show oscillations with periods of 3--5 min. We repeated the wavelet analysis for several neighboring regions of equal size outside the identified brightening. These regions also show comparable power in the 3-5 min range, indicating that the oscillatory behavior is not confined to the Ca II H brightening exceeding the threshold. However, the significant power does not always occur at exactly the same times, and cross-wavelet analysis shows that the neighboring signals are not uniformly phase-locked. We therefore interpret the oscillations as part of a spatially extended but locally structured and intermittent chromospheric oscillatory field.

The short-lived small-scale Ca~{\sc ii}~H\&K brightenings in the internetwork region are typically called Ca~{\sc ii}~H\&K grains \citep{Beckersphd}, which are widely believed to be the manifestations of acoustic shock heating happening when acoustic waves propagate upward in the solar atmosphere \citep{Carlsson1997,Wedemeyer2004,Beck2008,Mathur2022}. Previous papers suggest that the Ca~{\sc ii}~H\&K bright grains are correlated with the 3-min (5 mHz) intensity and velocity oscillations in the chromosphere \citep{Rutten1991, Kamio2006}. The good agreement between the characteristics of the Ca~{\sc ii}~H brightenings introduced in section~\ref{subsection:brightenings_weak_field}, and the general properties of the bright grains described above, supports the role of acoustic shock heating in producing these brightenings, a predominantly non-magnetic process.

Nonetheless, we do find small-scale photospheric magnetic features associated with the vast majority of these Ca~{\sc ii}~H brightenings. Also, over the evolution of the Ca~{\sc ii}~H brightenings over multiple periods, the peak intensities always coincide with peaks of the total unsigned magnetic flux in the corresponding region. These Ca~{\sc ii}~H brightenings could be the manifestation of magnetohydrodynamic waves traveling upward along the weak, small-scale, and short-lived magnetic fields \citep{Jafarzadeh2017a, MartinezGonzalez2023}. In this scenario the magnetic field serves mainly as a wave guide. Alternatively, small-scale photospheric magnetic cancellation/disappearance and emergence/appearance events are found to be associated with these brightenings, where the magnetic reconnection could potentially contribute to the excess chromospheric or coronal emission \citep{Gosic2018,Priest2018,Chitta2023}. In this picture, the magnetic field plays a much more active role in driving the chromospheric brightness.

In order to quantify the energy contribution from the magnetic field toward producing the brightenings, we compare the available magnetic energy and the energy required to heat the upper atmosphere. Following \cite{Kaithakkal2019}, we made a simple estimation for the case shown in Fig.~\ref{fig:case_class2_2}, where the cancellation between two opposite-polarity magnetic patches can be clearly seen. Since we do not have the temperature information at the height sampled by the {\sc Sunrise} Ca~{\sc ii}~H bandpass, we initially use the temperature at $\log\tau = -2.5$ to make a preliminary estimate.

(a) The total magnetic energy associated with the pair of canceling magnetic features can be calculated as $\delta E_{\mathrm{B}} = 2(B^2_{\mathrm{max}}/8\pi)A_{B,\mathrm{max}}h$, where $B_{\mathrm{max}} \approx 100 \ \mathrm{G}$ and the area at the time of maximum coverage, $A_{B,\mathrm{max}} \approx 6.6\times10^{14} \ \mathrm{cm^2}$ (50 pixels at IMaX plate scale) are respectively the maximum line-of-sight magnetic field strength and area of the positive-polarity patch, which undergoes complete cancellation with the negative polarity patch. The value $h\approx200 \ \mathrm{km}$ is roughly the height range between the layer where $\log\tau = -2.5$ and the IMaX magnetogram over which the heating takes place, and the factor of two accounts for the total energy from both polarities, assuming that the flux lost from the neighborhood positive-polarity magnetic patch is the same as that from the negative-polarity patch. In this case, we get $\delta E_{B}\approx1.05\times10^{25} \ \mathrm{erg}$. Note that we have assumed that the magnetic feature is fully resolved, i.e. that the measured magnetic field strength is the true value. For unresolved fields the area $A_{B,\mathrm{max}}$ decreases as $1/B$, while the energy increases as $B^2$. This implies that the above value is a lower limit of the true available magnetic energy.

(b) The thermal energy needed for the heating is estimated as $\delta E_{\mathrm{th}}=3/2nk_{\mathrm{B}}\delta TAh$, where $n, k_{\mathrm{B}}, \delta T, A$ are the particle number density at $\log\tau=-2.5$, the Boltzmann constant, the variation in temperature, and the area of the brightening. With $n\approx10^{16}\,\mathrm{cm^{-3}}$, $\delta T\approx200\,\mathrm{K}$, $A\approx1.3\times10^{15} \ \mathrm{cm^2}$ (100 pixels at the IMaX plate scale), we get $\delta E_{\mathrm{th}}\approx1.08\times10^{25} \ \mathrm{erg}$.

We can see that the magnetic energy $\delta E_{\mathrm{B}}$ is comparable with the thermal energy needed for heating, $\delta E_{\mathrm{th}}$. But the actual amount of magnetic energy that is released will be smaller than this value.  If we assume that 10\% of the magnetic energy is free energy, then it will be an  order of magnitude less than $\delta E_{\mathrm{th}}$. Furthermore, here we don't consider the radiated energy during the temperature enhancement, which could be even larger than the needed thermal energy in a small-scale magnetic cancellation event \citep{Kaithakkal2019}. Therefore, the magnetic reconnection between multiple internetwork magnetic patches by itself cannot support the enhancement of the temperature at $\log\tau=-2.5$, nor can it account for the energetics of these brightenings in the lower chromosphere. We propose that the (magneto-)acoustic shock heating is still the main cause of these Ca~{\sc ii}~H brightenings, while small-scale magnetic activity makes an additional contribution to many of them. This conclusion is supported by the fact that although Ca~{\sc ii}~H brightenings occur roughly periodically, particularly strong such brightenings generally coincide with emerging or canceling magnetic fields.

\citet{Schrijver1989} mentioned the relatively strong Ca {\sc ii} K emission with a relatively small magnetic flux density in an active region, which was mainly discovered in the immediate surroundings of sunspots, or the plage regions away from sunspots. They speculated that the inclination of magnetic flux tubes and unresolved opposite-polarity magnetic features could be responsible. Our analysis gives credence to these two factors, although there might be some differences between the events seen in active region and those that occur in the quiet Sun.

\section{Summary and conclusions}
\label{section:conclusion}

In this paper, we use high resolution data acquired by the {\sc Sunrise} balloon-borne observatory during its first science flight in June 2009, to study the nature of the Ca~{\sc ii}~H brightenings associated with weak fields in the quiet Sun. Roughly half of these brightenings are located at peripheries of strong field regions (often belonging to the magnetic network). We found that they are likely associated with the magnetic canopy overlying weak or no field, which are produced by the expansion of magnetic features with height. A similar effect occurs when the magnetic element is inclined with respect to the solar surface normal.

Other Ca~{\sc ii}~H brightenings are found in regions with exclusively weak fields as seen by {\sc Sunrise}/IMaX. We categorize them into three classes based on their associated magnetic features. $Class \ 1$: brightenings in unipolar magnetic patches. These are likely spatially unresolved magnetic bright points. I.e. they are finally also associated with strong magnetic fields. $Class \ 2$: brightenings associated with the interaction between mixed-polarity magnetic patches. These brightenings show a roughly periodic behavior in the 3-5 min range. They share many properties with Ca~{\sc ii}~H and K grains. The association of interacting magnetic features with particularly strong brightenings suggests that the magnetic field does contribute to the brightening (e.g. through heating via magnetic reconnection), although (magneto-)acoustic waves play a strong role; $Class \ 3$: brightenings that are not associated with any significant magnetic fields. These are otherwise very similar to the brightenings in Class 2. We cannot rule out the presence of very weak, unresolved magnetic features. Only a very small fraction of all identified Ca~{\sc ii}~H brightenings belong to this class. However, we note that we have only studied the strongest brightenings associated with particularly weak fields. There are many more brightenings that are not so strong, but still well above the average Ca~{\sc ii}~H brightness in pixels with weak or no fields. Since SuFI provided no detailed spectral information, we cannot check if some of these aren't more normal Ca~{\sc ii}~H and K grains. 

These results contribute to our understanding of the nature of chromospheric brightenings in quiet Sun regions, especially the short-lived brightenings that are widely recognized as Ca grains. From the high-resolution observations of the photospheric magnetic field, we found flux emergence and cancellation between weak opposite polarity magnetic patches that are associated with the chromospheric grains. These various magnetic interactions could contribute to the Ca~{\sc ii}~H brightenings in addition to (magneto-)acoustic shock heating in the weak field region.

Further research is planned to improve our results based on the findings obtained so far. We plan to reproduce these Ca~{\sc ii}~H brightenings using MHD simulations, to better understand their magnetic nature, as well as the role that the magnetic field plays in the periodic variations in the Ca~{\sc ii}~H intensity. Moreover, this study will be extended to the high spatial and temporal resolution data from {\sc Sunrise} III \citep{Lagg2025, Solanki2026}, which was launched successfully in July 2024. The three instruments on {\sc Sunrise} III, SCIP \citep{Katsukawa2026}, SUSI \citep{Feller2025,Iglesias2025}, and TuMag \citep{delToro2025} record the full Stokes vector at photospheric and chromospheric heights. In particular, slit spectropolarimeter SUSI has observed Ca~{\sc ii}~H and K line profiles in all four Stokes parameters. This will allow robustly identifying Ca~{\sc ii}~H$_{2v}$ and/or K$_{2v}$ grains, thus filling an important gap in the observations analyzed here.

\begin{acknowledgements}
      This project has received funding from the European Research Council (ERC) under the European Union's Horizon 2020 research and innovation programme (grant agreement No. 101097844 — project WINSUN - and grant agreement No. 101039844 - project ORIGIN). Views and opinions expressed are however those of the author(s) only and do not necessarily reflect those of the European Union or the European Research Council. Neither the European Union nor the granting authority can be held responsible for them.
      The German contribution to {\sc Sunrise} is funded by the Bundesministerium f\"{u}r Wirtschaft und Technologie through the Deutsches Zentrum f\"{u}r Luft- und Raumfahrt e.V. (DLR), grant No. 50 OU 0401, and by the Innovationsfond of the President of the Max Planck Society (MPG). The Spanish contribution has been funded by the Spanish MICINN under projects ESP2006-13030-C06 and AYA2009-14105-C06 (including European FEDER funds). The HAO contribution was partly funded through NASA grant NNX08AH38G. 
      SJ acknowledges support from the UK Science and Technology Facilities Council (STFC) through consolidated grants ST/T00021X/1 and ST/X000923/1.
\end{acknowledgements}

\bibliography{reference}

@article{Vernazza1981,
        author = {{Vernazza}, J.~E. and {Avrett}, E.~H. and {Loeser}, R.},
        title = "{Structure of the solar chromosphere. III. Models of the EUV brightness components of the quiet sun.}",
        journal   = {\apjs},
        year = 1981,
        volume = {45},
        pages = {635-725},
        language  = {english}
        }

@ARTICLE{Solanki2026,
       author = {{Solanki}, Sami K. and {Smitha}, H.~N. and {Lagg}, Andreas and {Gandorfer}, Achim and {del Toro Iniesta}, Jose Carlos and {Katsukawa}, Yukio and {Bernasconi}, Pietro and {Berkefeld}, Thomas and {Feller}, Alex and {Riethm{\"u}ller}, Tino L. and {{\'A}lvarez-Herrero}, Alberto and {Kubo}, Masahito and {Orozco Su{\'a}rez}, David and {Grauf}, Bianca and {Carpenter}, Michael and {Bell}, Alexander and {Mart{\'\i}nez Pillet}, Valent{\'\i}n and {Gizon}, Laurent and {Bail{\'e}n}, Francisco Javier and {Blanco Rodr{\'\i}guez}, Julian and {Castellanos Dur{\'a}n}, Juan Sebasti{\'a}n and {Harnes}, Edvarda and {Hoelken}, Johannes and {Iglesias}, Francisco A. and {Ishikawa}, Ryohtaroh T. and {Kawabata}, Yusuke and {Matsumoto}, Takuma and {Oba}, Takayoshi and {Singh}, Kunal H. and {Siu-Tapia}, Azaymi L. and {Strecker}, Hanna and {Vukadinovic}, Du{\v{s}}an and {van Noort}, Michiel and {Balaguer Jim{\'e}nez}, Maria and {Sanchis Kilders}, Esteban and {Torralbo}, Ignacio and {Kuckein}, Christoph and {Hara}, Hirohisa and {Shimizu}, Toshifumi and {Volkmer}, Reiner and {Preis}, Tobias and {Raouafi}, Nour E. and {Vourlidas}, Angelos and {Hirzberger}, Johann and {Deutsch}, Werner and {Germerott}, Dietmar and {Heerlein}, Klaus and {Kolleck}, Martin and {{\'A}lvarez Garc{\'\i}a}, Daniel and {L{\'o}pez Jim{\'e}nez}, Antonio C. and {Bellot Rubio}, Luis R. and {Morales-Fern{\'a}ndez}, Jos{\'e} Miguel and {Moreno Mantas}, Antonio Jes{\'u}s and {Aparicio del Moral}, Beatriz and {S{\'a}nchez G{\'o}mez}, Antonio and {Bail{\'o}n Mart{\'\i}nez}, Eduardo and {Santamarina Guerrero}, Pablo and {Hern{\'a}ndez Exp{\'o}sito}, David and {Tobaruela}, Angel and {Gasent Blesa}, Jos{\'e} Luis and {Schulze}, Erich and {Eaton}, Harry and {Palo}, Geoffrey and {Ayoub}, Daniel and {Naito}, Yoshihiro and {Quintero Noda}, Carlos and {Uraguchi}, Fumihiro and {Tsuzuki}, Toshihiro and {Piqueras Carre{\~n}o}, Javier},
        title = "{SUNRISE III: Instrument, Mission, Data, and First Results}",
      journal = {\apjl},
         year = 2026,
        month = jul,
       volume = {1005},
       number = {2},
          eid = {L64},
        pages = {L64},
          doi = {10.3847/2041-8213/ae796b},
archivePrefix = {arXiv},
       eprint = {2606.07989},
 primaryClass = {astro-ph.SR},
       adsurl = {https://ui.adsabs.harvard.edu/abs/2026ApJ..1005L..64S}
}

@article{Anusha2017,
        author = {{Anusha}, L.~S. and {Solanki}, S.~K. and {Hirzberger}, J. and {Feller}, A.},
        title = "{Statistical evolution of quiet-Sun small-scale magnetic features using Sunrise observations}",
        journal = {\aap},
        year = 2017,
        volume = {598},
        pages = {A47},
        language  = {english}
        }

@INPROCEEDINGS{Solanki2004,
        author = {{Solanki}, Sami K.},
        title = "{Structure of the solar chromosphere}",
        booktitle = {Multi-Wavelength Investigations of Solar Activity},
        year = 2004,
        editor = {{Stepanov}, Alexander V. and {Benevolenskaya}, Elena E. and {Kosovichev}, Alexander G.},
        volume = {223},
        pages = {195-202},
        language  = {english}
        }

@article{Carlsson2019,
        author = {{Carlsson}, Mats and {De Pontieu}, Bart and {Hansteen}, Viggo H.},
        title = "{New View of the Solar Chromosphere}",
        journal   = {\araa},
        year = 2019,
        volume = {57},
        pages = {189-226},
        language  = {english}
        }

@article{Carlsson1992,
        author = {{Carlsson}, Mats and {Stein}, Robert F.},
        title = "{Non-LTE Radiating Acoustic Shocks and CA II K2V Bright Points}",
        journal = {\apjl},
        year = 1992,
        volume = {397},
        pages = {L59},
        language  = {english}
        }

@article{Carlsson1997,
        author = {{Carlsson}, Mats and {Stein}, Robert F.},
        title = "{Formation of Solar Calcium H and K Bright Grains}",
        journal = {\apj},
        year = 1997,
        month = may,
        volume = {481},
        pages = {500-514},
        language  = {english}
        }

@article{Wedemeyer2004,
        author = {{Wedemeyer}, S. and {Freytag}, B. and {Steffen}, M. and {Ludwig}, H. -G. and {Holweger}, H.},
        title = "{Numerical simulation of the three-dimensional structure and dynamics of the non-magnetic solar chromosphere}",
        journal = {\aap},
        year = 2004,
        volume = {414},
        pages = {1121-1137},
        language  = {english}
        }

@phdthesis{Beckersphd,
        author = {{Beckers}, Jacques Maurice},
        title = "{A study of the fine structure in the solar chromosphere}",
        school = {University of Utrecht, Netherlands},
        year      = {1964},
        language  = {english}
        }

@article{Mathur2022,
        author = {{Mathur}, Harsh and {Joshi}, Jayant and {Nagaraju}, K. and {Rouppe van der Voort}, Luc and {Bose}, Souvik},
        title = "{Properties of shock waves in the quiet-Sun chromosphere}",
        journal = {\aap},
        year = 2022,
        volume = {668},
        pages = {A153},
        language  = {english}
        }

@article{Jafarzadeh2017a,
        author = {{Jafarzadeh}, S. and {Solanki}, S.~K. and {Stangalini}, M. and {Steiner}, O. and {Cameron}, R.~H. and {Danilovic}, S.},
        title = "{High-frequency Oscillations in Small Magnetic Elements Observed with Sunrise/SuFI}",
        journal = {\apjs},
        year = 2017,
        volume = {229},
        pages = {10},
        language  = {english}
        }

@article{Gosic2018,
        author    = {{Go{\v{s}}i{\'c}}, M. and {de la Cruz Rodr{\'\i}guez}, J. and {De Pontieu}, B. and {Bellot Rubio}, L.~R. and {Carlsson}, M. and {Esteban Pozuelo}, S. and {Ortiz}, A. and {Polito}, V.},
        title     = {Chromospheric Heating due to Cancellation of Quiet Sun Internetwork Fields},
        journal   = {\apj},
        volume    = {857},
        pages   = {48},
        year      = {2018},
        language  = {english}
        }

@article{Rutten1991,
        author = {{Rutten}, Robert J. and {Uitenbroek}, Han},
        title = "{Ca {\scshape ii} H$_{2v}$ and K$_{2v}$ cell grains}",
        journal = {\solphys},
        year = 1991,
        volume = {134},
        pages = {15-71},
        language  = {english}
}

@article{Schrijver1989,
        author = {{Schrijver}, C.~J. and {Cote}, J. and {Zwaan}, C. and {Saar}, S.~H.},
        title = "{Relations between the Photospheric Magnetic Field and the Emission from the Outer Atmospheres of Cool Stars. I. The Solar CA II K Line Core Emission}",
        journal = {\apj},
        year = 1989,
        volume = {337},
        pages = {964},
        language  = {english}
}

@article{Harvey1999,
        author = {{Harvey}, Karen L. and {White}, Oran R.},
        title = "{Magnetic and Radiative Variability of Solar Surface Structures. I. Image Decomposition and Magnetic-Intensity Mapping}",
        journal = {\apj},
        year = 1999,
        volume = {515},
        pages = {812-831},
        language  = {english}
}

@article{Ortiz2005,
        author = {{Ortiz}, A. and {Rast}, M.},
        title = "{How good is the Ca II K as a proxy for the magnetic flux?}",
        journal = {Memorie della Societ$\grave{a}$ Astronomica Italiana},
        year = 2005,
        volume = {76},
        pages = {1018},
        language  = {english}
}

@article{Loukitcheva2009,
        author = {{Loukitcheva}, M. and {Solanki}, S.~K. and {White}, S.~M.},
        title = "{The relationship between chromospheric emissions and magnetic field strength}",
        journal = {\aap},
        year = 2009,
        volume = {497},
        pages = {273-285},
        language  = {english}
}

@article{Barthol2011,
        author = {{Barthol}, P. and {Gandorfer}, A. and {Solanki}, S.~K. and {Sch{\"u}ssler}, M. and {Chares}, B. and {Curdt}, W. and {Deutsch}, W. and {Feller}, A. and {Germerott}, D. and {Grauf}, B. and {Heerlein}, K. and {Hirzberger}, J. and {Kolleck}, M. and {Meller}, R. and {M{\"u}ller}, R. and {Riethm{\"u}ller}, T.~L. and {Tomasch}, G. and {Kn{\"o}lker}, M. and {Lites}, B.~W. and {Card}, G. and {Elmore}, D. and {Fox}, J. and {Lecinski}, A. and {Nelson}, P. and {Summers}, R. and {Watt}, A. and {Mart{\'\i}nez Pillet}, V. and {Bonet}, J.~A. and {Schmidt}, W. and {Berkefeld}, T. and {Title}, A.~M. and {Domingo}, V. and {Gasent Blesa}, J.~L. and {del Toro Iniesta}, J.~C. and {L{\'o}pez Jim{\'e}nez}, A. and {{\'A}lvarez-Herrero}, A. and {Sabau-Graziati}, L. and {Widani}, C. and {Haberler}, P. and {H{\"a}rtel}, K. and {Kampf}, D. and {Levin}, T. and {P{\'e}rez Grande}, I. and {Sanz-Andr{\'e}s}, A. and {Schmidt}, E.},
        title = "{The Sunrise Mission}",
        journal = {\solphys},
        year = 2011,
        volume = {268},
        number = {1},
        pages = {1-34},
        language  = {english}
        }

@article{Solanki2010,
        author = {{Solanki}, S.~K. and {Barthol}, P. and {Danilovic}, S. and {Feller}, A. and {Gandorfer}, A. and {Hirzberger}, J. and {Riethm{\"u}ller}, T.~L. and {Sch{\"u}ssler}, M. and {Bonet}, J.~A. and {Mart{\'\i}nez Pillet}, V. and {del Toro Iniesta}, J.~C. and {Domingo}, V. and {Palacios}, J. and {Kn{\"o}lker}, M. and {Bello Gonz{\'a}lez}, N. and {Berkefeld}, T. and {Franz}, M. and {Schmidt}, W. and {Title}, A.~M.},
        title = "{SUNRISE: Instrument, Mission, Data, and First Results}",
        journal = {\apjl},
        year = 2010,
        volume = {723},
        pages = {L127-L133},
        language  = {english}
        }

@article{Gandorfer2011,
        author = {{Gandorfer}, A. and {Grauf}, B. and {Barthol}, P. and {Riethm{\"u}ller}, T.~L. and {Solanki}, S.~K. and {Chares}, B. and {Deutsch}, W. and {Ebert}, S. and {Feller}, A. and {Germerott}, D. and {Heerlein}, K. and {Heinrichs}, J. and {Hirche}, D. and {Hirzberger}, J. and {Kolleck}, M. and {Meller}, R. and {M{\"u}ller}, R. and {Sch{\"a}fer}, R. and {Tomasch}, G. and {Kn{\"o}lker}, M. and {Mart{\'\i}nez Pillet}, V. and {Bonet}, J.~A. and {Schmidt}, W. and {Berkefeld}, T. and {Feger}, B. and {Heidecke}, F. and {Soltau}, D. and {Tischenberg}, A. and {Fischer}, A. and {Title}, A. and {Anwand}, H. and {Schmidt}, E.},
        title = "{The Filter Imager SuFI and the Image Stabilization and Light Distribution System ISLiD of the Sunrise Balloon-Borne Observatory: Instrument Description}",
        journal = {\solphys},
        year = 2011,
        volume = {268},
        pages = {35-55},
        language  = {english}
        }

@article{Martinez2011,
        author = {{Mart{\'\i}nez Pillet}, V. and {del Toro Iniesta}, J.~C. and {{\'A}lvarez-Herrero}, A. and {Domingo}, V. and {Bonet}, J.~A. and {Gonz{\'a}lez Fern{\'a}ndez}, L. and {L{\'o}pez Jim{\'e}nez}, A. and {Pastor}, C. and {Gasent Blesa}, J.~L. and {Mellado}, P. and {Piqueras}, J. and {Aparicio}, B. and {Balaguer}, M. and {Ballesteros}, E. and {Belenguer}, T. and {Bellot Rubio}, L.~R. and {Berkefeld}, T. and {Collados}, M. and {Deutsch}, W. and {Feller}, A. and {Girela}, F. and {Grauf}, B. and {Heredero}, R.~L. and {Herranz}, M. and {Jer{\'o}nimo}, J.~M. and {Laguna}, H. and {Meller}, R. and {Men{\'e}ndez}, M. and {Morales}, R. and {Orozco Su{\'a}rez}, D. and {Ramos}, G. and {Reina}, M. and {Ramos}, J.~L. and {Rodr{\'\i}guez}, P. and {S{\'a}nchez}, A. and {Uribe-Patarroyo}, N. and {Barthol}, P. and {Gandorfer}, A. and {Knoelker}, M. and {Schmidt}, W. and {Solanki}, S.~K. and {Vargas Dom{\'\i}nguez}, S.},
        title = "{The Imaging Magnetograph eXperiment (IMaX) for the Sunrise Balloon-Borne Solar Observatory}",
        journal = {\solphys},
        year = 2011,
        volume = {268},
        pages = {57-102},
        language  = {english}
        }

@phdthesis{Solankiphd,
        author    = {Solanki, S. K.},
        title     = "{The Photospheric Layers of Solar Magnetic Fluxfubes}",
        school   = {ETH Z\"urich},
        year      = {1987},
        language  = {english}
        }

@phdthesis{Frutigerphd,
        author    = {Frutiger, C.},
        title     = {},
        school   = {ETH Z\"urich},
        year      = {2000},
        language  = {english}
        }

@article{Frutiger2000,
        author = {{Frutiger}, C. and {Solanki}, S.~K. and {Fligge}, M. and {Bruls}, J.~H.~M.~J.},
        title = "{Properties of the solar granulation obtained from the inversion of low spatial resolution spectra}",
        journal = {\aap},
        year = 2000,
        volume = {358},
        pages = {1109-1121},
        language  = {english}
        }

@article{Kahil2017,
        author = {{Kahil}, F. and {Riethm{\"u}ller}, T.~L. and {Solanki}, S.~K.},
        title = "{Brightness of Solar Magnetic Elements As a Function of Magnetic Flux at High Spatial Resolution}",
        journal   = {\apjs},
        volume    = {229},
        pages   = {12},
        year      = {2017},
        language  = {english}
        }

@article{Riethmuller2017,
        author = {{Riethm{\"u}ller}, T.~L. and {Solanki}, S.~K. and {Barthol}, P. and {Gandorfer}, A. and {Gizon}, L. and {Hirzberger}, J. and {van Noort}, M. and {Blanco Rodr{\'\i}guez}, J. and {Del Toro Iniesta}, J.~C. and {Orozco Su{\'a}rez}, D. and {Schmidt}, W. and {Mart{\'\i}nez Pillet}, V. and {Kn{\"o}lker}, M.},
        title = "{A New MHD-assisted Stokes Inversion Technique}",
        journal   = {\apjs},
        volume    = {229},
        pages   = {16},
        year      = {2017},
        language  = {english}
        }

@article{Jafarzadeh2014,
        author = {{Jafarzadeh}, S. and {Solanki}, S.~K. and {Lagg}, A. and {Bellot Rubio}, L.~R. and {van Noort}, M. and {Feller}, A. and {Danilovic}, S.},
        title = "{Inclinations of small quiet-Sun magnetic features based on a new geometric approach}",
        journal   = {\aap},
        volume    = {569},
        pages   = {A105},
        year      = {2014},
        language  = {english}
        }

@article{Nagata2008,
        author    = {{Nagata}, Shin'ichi and {Tsuneta}, Saku and {Suematsu}, Yoshinori and {Ichimoto}, Kiyoshi and {Katsukawa}, Yukio and {Shimizu}, Toshifumi and {Yokoyama}, Takaaki and {Tarbell}, Theodore D. and {Lites}, Bruce W. and {Shine}, Richard A. and {Berger}, Thomas E. and {Title}, Alan M. and {Bellot Rubio}, Luis R. and {Orozco Su{\'a}rez}, David},
        title     = {Formation of Solar Magnetic Flux Tubes with Kilogauss Field Strength Induced by Convective Instability},
        journal   = {\apjl},
        volume = {677},
        pages = {L145},
        year      = {2008},
        language  = {english}
        }

@article{Riethmuller2014,
        author = {{Riethm{\"u}ller}, T.~L. and {Solanki}, S.~K. and {Berdyugina}, S.~V. and {Sch{\"u}ssler}, M. and {Mart{\'\i}nez Pillet}, V. and {Feller}, A. and {Gandorfer}, A. and {Hirzberger}, J.},
        title = "{Comparison of solar photospheric bright points between Sunrise observations and MHD simulations}",
        journal   = {\aap},
        volume    = {568},
        pages   = {A13},
        year      = {2014},
        language  = {english}
        }

@article{Jafarzadeh2013,
        author    = {{Jafarzadeh}, S. and {Solanki}, S.~K. and {Feller}, A. and {Lagg}, A. and {Pietarila}, A. and {Danilovic}, S. and {Riethm{\"u}ller}, T.~L. and {Mart{\'\i}nez Pillet}, V.},
        title     = {Structure and dynamics of isolated internetwork Ca II H bright points observed by SUNRISE},
        journal   = {\aap},
        volume    = {549},
        pages   = {A116},
        year      = {2013},
        language  = {english}
        }

@article{Riethmuller2010,
       author = {{Riethm{\"u}ller}, T.~L. and {Solanki}, S.~K. and {Mart{\'\i}nez Pillet}, V. and {Hirzberger}, J. and {Feller}, A. and {Bonet}, J.~A. and {Bello Gonz{\'a}lez}, N. and {Franz}, M. and {Sch{\"u}ssler}, M. and {Barthol}, P. and {Berkefeld}, T. and {del Toro Iniesta}, J.~C. and {Domingo}, V. and {Gandorfer}, A. and {Kn{\"o}lker}, M. and {Schmidt}, W.},
        title = "{Bright Points in the Quiet Sun as Observed in the Visible and Near-UV by the Balloon-borne Observatory SUNRISE}",
        journal   = {\apjl},
        volume = {723},
        pages = {L169-L174},
        year      = {2010},
        language  = {english}
}

@article{Spruit1976,
       author = {{Spruit}, H.~C.},
        title = "{Pressure equilibrium and energy balance of small photospheric fluxtubes}",
        journal   = {\solphys},
        volume = {50},
        pages = {269-295},
        year      = {1976},
        language  = {english}
}

@article{Beck2008,
        author = {{Beck}, C. and {Schmidt}, W. and {Rezaei}, R. and {Rammacher}, W.},
        title = "{The signature of chromospheric heating in Ca II H spectra}",
        journal = {\aap},
        year = 2008,
        volume = {479},
        pages = {213-227},
        language  = {english}
}

@article{Lites1999,
        author = {{Lites}, B.~W. and {Rutten}, R.~J. and {Berger}, T.~E.},
        title = "{Dynamics of the Solar Chromosphere. II. Ca II H$_{2V}$ and K$_{2V}$ Grains versus Internetwork Fields}",
        journal = {\apj},
        year = 1999,
        volume = {517},
        pages = {1013-1033},
        language  = {english}
}

@article{Kamio2006,
        author = {S. Kamio and H. Kurokawa},
        title = "{The relation between Ca bright grains and oscillations in the photosphere}",
        journal = {\aap},
        year = 2006,
        volume = {450},
        pages = {351-358},
        language  = {english}
}

@article{Lin1995,
        author = {{Lin}, Haosheng},
        title = "{On the Distribution of the Solar Magnetic Fields}",
        journal = {\apj},
        year = 1995,
        volume = {446},
        pages = {421},
        language  = {english}
}

@article{Khomenko2003,
        author = {{Khomenko}, E.~V. and {Collados}, M. and {Solanki}, S.~K. and {Lagg}, A. and {Trujillo Bueno}, J.},
        title = "{Quiet-Sun inter-network magnetic fields  observed in the infrared}",
        journal = {\aap},
        year = 2003,
        volume = {408},
        pages = {1115-1135},
        language  = {english}
}

@article{Orozco2007,
        author = {{Orozco Su{\'a}rez}, D. and {Bellot Rubio}, L.~R. and {del Toro Iniesta}, J.~C. and {Tsuneta}, S. and {Lites}, B.~W. and {Ichimoto}, K. and {Katsukawa}, Y. and {Nagata}, S. and {Shimizu}, T. and {Shine}, R.~A. and {Suematsu}, Y. and {Tarbell}, T.~D. and {Title}, A.~M.},
        title = "{Quiet-Sun Internetwork Magnetic Fields from the Inversion of Hinode Measurements}",
        journal = {\apj},
        year = 2007,
        volume = {670},
        pages = {L61-L64},
        language  = {english}
}

@article{Martinez2008,
        author = {{Mart{\'\i}nez Gonz{\'a}lez}, M.~J. and {Collados}, M. and {Ruiz Cobo}, B. and {Beck}, C.},
        title = "{Internetwork magnetic field distribution from simultaneous 1.56 {\ensuremath{\mu}}m and 630 nm observations}",
        journal = {\aap},
        year = 2008,
        volume = {477},
        pages = {953-965},
        language  = {english}
}

@article{GrossmannDoerth1988,
        author = {{Grossmann-Doerth}, U. and {Sch{\"u}ssler}, M. and {Solanki}, S.~K.},
        title = "{Unshifted, asymmetric Stokes V-profiles - Possible solution of a riddle}",
        journal = {\aap},
        year = 1988,
        volume = {206},
        pages = {L37-L39},
        language  = {english}
}

@article{Solanki1989,
        author = {{Solanki}, S.~K.},
        title = "{The origin and the diagnostic capabilities of the Stokes V asymmetry observed in solar faculae and the network}",
        journal = {\aap},
        year = 1989,
        volume = {224},
        pages = {225-241},
        language  = {english}
}

@article{Buehler2015,
        author = {{Buehler}, D. and {Lagg}, A. and {Solanki}, S.~K. and {van Noort}, M.},
        title = "{Properties of solar plage from a spatially coupled inversion of Hinode SP data}",
        journal = {\aap},
        year = 2015,
        volume = {576},
        pages = {A27},
        language  = {english}
}

@article{Kaithakkal2019,
        author = {{Kaithakkal}, A.~J. and {Solanki}, S.~K.},
        title = "{Cancelation of small-scale magnetic features}",
        journal = {\aap},
        year = 2019,
        volume = {622},
        pages = {A200},
        language  = {english}
}

@article{Berkefeld2011,
        author = {{Berkefeld}, T. and {Schmidt}, W. and {Soltau}, D. and {Bell}, A. and {Doerr}, H.~P. and {Feger}, B. and {Friedlein}, R. and {Gerber}, K. and {Heidecke}, F. and {Kentischer}, T. and {v. d. L{\"u}he}, O. and {Sigwarth}, M. and {W{\"a}lde}, E. and {Barthol}, P. and {Deutsch}, W. and {Gandorfer}, A. and {Germerott}, D. and {Grauf}, B. and {Meller}, R. and {{\'A}lvarez-Herrero}, A. and {Kn{\"o}lker}, M. and {Mart{\'\i}nez Pillet}, V. and {Solanki}, S.~K. and {Title}, A.~M.},
        title = "{The Wave-Front Correction System for the Sunrise Balloon-Borne Solar Observatory}",
        journal = {\solphys},
        year = 2011,
        volume = {268},
        pages = {103-123},
        language  = {english}
}

@article{Hirzberger2010,
        author = {{Hirzberger}, J. and {Feller}, A. and {Riethm{\"u}ller}, T.~L. and {Sch{\"u}ssler}, M. and {Borrero}, J.~M. and {Afram}, N. and {Unruh}, Y.~C. and {Berdyugina}, S.~V. and {Gandorfer}, A. and {Solanki}, S.~K. and {Barthol}, P. and {Bonet}, J.~A. and {Mart{\'\i}nez Pillet}, V. and {Berkefeld}, T. and {Kn{\"o}lker}, M. and {Schmidt}, W. and {Title}, A.~M.},
        title = "{Quiet-sun Intensity Contrasts in the Near-ultraviolet as Measured from SUNRISE}",
        journal = {\apjl},
        year = 2010,
        volume = {723},
        pages = {L154-L158}
}

@article{Hirzberger2011,
        author = {{Hirzberger}, J. and {Feller}, A. and {Riethm{\"u}ller}, T.~L. and {Gandorfer}, A. and {Solanki}, S.~K.},
        title = "{Performance validation of phase diversity image reconstruction techniques}",
        journal = {\aap},
        year = 2011,
        volume = {529},
        pages = {A132},
}

@article{Solanki2017,
        author = {{Solanki}, S.~K. and {Riethm{\"u}ller}, T.~L. and {Barthol}, P. and {Danilovic}, S. and {Deutsch}, W. and {Doerr}, H. -P. and {Feller}, A. and {Gandorfer}, A. and {Germerott}, D. and {Gizon}, L. and {Grauf}, B. and {Heerlein}, K. and {Hirzberger}, J. and {Kolleck}, M. and {Lagg}, A. and {Meller}, R. and {Tomasch}, G. and {van Noort}, M. and {Blanco Rodr{\'\i}guez}, J. and {Gasent Blesa}, J.~L. and {Balaguer Jim{\'e}nez}, M. and {Del Toro Iniesta}, J.~C. and {L{\'o}pez Jim{\'e}nez}, A.~C. and {Orozco Suarez}, D. and {Berkefeld}, T. and {Halbgewachs}, C. and {Schmidt}, W. and {{\'A}lvarez-Herrero}, A. and {Sabau-Graziati}, L. and {P{\'e}rez Grande}, I. and {Mart{\'\i}nez Pillet}, V. and {Card}, G. and {Centeno}, R. and {Kn{\"o}lker}, M. and {Lecinski}, A.},
        title = "{The Second Flight of the Sunrise Balloon-borne Solar Observatory: Overview of Instrument Updates, the Flight, the Data, and First Results}",
        journal = {\apjs},
        year = 2017,
        volume = {229},
        pages = {2},
}

@article{Cram1977,
        author = {{Cram}, L.~E. and {Brown}, D.~R. and {Beckers}, J.~M.},
        title = "{High resolution spectroscopy of the disk chromosphere. V. Space-time variations observed simultaneously in seven lines.}",
        journal = {\aap},
        year = 1977,
        volume = {57},
        pages = {211-220},
}

@article{Remling1996,
        author = {{Remling}, B. and {Deubner}, F. -L. and {Steffens}, S.},
        title = "{Evidence of K\_2v\_ grains being a non-magnetic phenomenon.}",
        journal = {\aap},
        year = 1996,
        volume = {316},
        pages = {196-200},
        adsurl = {https://ui.adsabs.harvard.edu/abs/1996A&A...316..196R}
}

@article{Sivaraman1982,
        author = {{Sivaraman}, K.~R. and {Livingston}, W.~C.},
        title = "{Ca  ii K$_{2V}$ spectral features and their relation to small-scale photospheric magnetic fields}",
        journal = {\solphys},
        year = 1982,
        volume = {80},
        pages = {227-231},
}

@article{Sivaraman2000,
        author = {{Sivaraman}, K.~R. and {Gupta}, S.~S. and {Livingston}, W.~C. and {Dam{\'e}}, L. and {Kalkofen}, W. and {Keller}, C.~U. and {Smartt}, R. and {Hasan}, S.~S.},
        title = "{Results from a revisit to the K$_{2V}$ bright points}",
        journal = {\aap},
        year = 2000,
        volume = {363},
        pages = {279-288},
}

@article{MartinezGonzalez2023,
        author = {{Mart{\'\i}nez Gonz{\'a}lez}, Mar{\'\i}a Jes{\'u}s and {del Pino Alem{\'a}n}, Tanaus{\'u} and {Yabar}, Adur Pastor and {Noda}, Carlos Quintero and {Ramos}, Andr{\'e}s Asensio},
        title = "{On the Magnetic Nature of Quiet-Sun Chromospheric Grains}",
        journal = {\apjl},
        year = 2023,
        volume = {955},
        pages = {L40},
}

@article{Worden1999,
        author = {{Worden}, John and {Harvey}, John and {Shine}, Richard},
        title = "{Bright Chromospheric Grains and the Magnetic Intranetwork}",
        journal = {\apj},
        year = 1999,
        volume = {523},
        pages = {450-457},
}

@ARTICLE{Kneer1993,
       author = {{Kneer}, F. and {von Uexkuell}, M.},
        title = "{Oscillations of the Sun's chromosphere. VI. K grains, resonances, and gravity waves.}",
      journal = {\aap},
         year = 1993,
        month = jul,
       volume = {274},
        pages = {584-594},
       adsurl = {https://ui.adsabs.harvard.edu/abs/1993A&A...274..584K}
}

@ARTICLE{Chitta2023,
       author = {{Chitta}, L.~P. and {Solanki}, S.~K. and {del Toro Iniesta}, J.~C. and {Woch}, J. and {Calchetti}, D. and {Gandorfer}, A. and {Hirzberger}, J. and {Kahil}, F. and {Valori}, G. and {Orozco Su{\'a}rez}, D. and {Strecker}, H. and {Appourchaux}, T. and {Volkmer}, R. and {Peter}, H. and {Mandal}, S. and {Aznar Cuadrado}, R. and {Teriaca}, L. and {Sch{\"u}hle}, U. and {Berghmans}, D. and {Verbeeck}, C. and {Zhukov}, A.~N. and {Priest}, E.~R.},
        title = "{Fleeting Small-scale Surface Magnetic Fields Build the Quiet-Sun Corona}",
      journal = {\apjl},
         year = 2023,
        month = oct,
       volume = {956},
       number = {1},
          eid = {L1},
        pages = {L1},
          doi = {10.3847/2041-8213/acf136},
archivePrefix = {arXiv},
       eprint = {2308.10982},
 primaryClass = {astro-ph.SR},
       adsurl = {https://ui.adsabs.harvard.edu/abs/2023ApJ...956L...1C}
}

@ARTICLE{Priest2018,
       author = {{Priest}, E.~R. and {Chitta}, L.~P. and {Syntelis}, P.},
        title = "{A Cancellation Nanoflare Model for Solar Chromospheric and Coronal Heating}",
      journal = {\apjl},
         year = 2018,
        month = aug,
       volume = {862},
       number = {2},
          eid = {L24},
        pages = {L24},
          doi = {10.3847/2041-8213/aad4fc},
archivePrefix = {arXiv},
       eprint = {1807.08161},
 primaryClass = {astro-ph.SR},
       adsurl = {https://ui.adsabs.harvard.edu/abs/2018ApJ...862L..24P}
}

@ARTICLE{Lagg2010,
       author = {{Lagg}, A. and {Solanki}, S.~K. and {Riethm{\"u}ller}, T.~L. and {Mart{\'\i}nez Pillet}, V. and {Sch{\"u}ssler}, M. and {Hirzberger}, J. and {Feller}, A. and {Borrero}, J.~M. and {Schmidt}, W. and {del Toro Iniesta}, J.~C. and {Bonet}, J.~A. and {Barthol}, P. and {Berkefeld}, T. and {Domingo}, V. and {Gandorfer}, A. and {Kn{\"o}lker}, M. and {Title}, A.~M.},
        title = "{Fully Resolved Quiet-Sun Magnetic flux Tube Observed with the SUNRISE/IMAX Instrument}",
      journal = {\apjl},
         year = 2010,
        month = nov,
       volume = {723},
       number = {2},
        pages = {L164-L168},
          doi = {10.1088/2041-8205/723/2/L164},
archivePrefix = {arXiv},
       eprint = {1009.0996},
 primaryClass = {astro-ph.SR},
       adsurl = {https://ui.adsabs.harvard.edu/abs/2010ApJ...723L.164L}
}

@ARTICLE{Bhatia2022,
       author = {{Bhatia}, Tanayveer S. and {Cameron}, Robert H. and {Solanki}, Sami K. and {Peter}, Hardi and {Przybylski}, Damien and {Witzke}, Veronika and {Shapiro}, Alexander},
        title = "{Small-scale dynamo in cool stars. I. Changes in stratification and near-surface convection for main-sequence spectral types}",
      journal = {\aap},
         year = 2022,
        month = jul,
       volume = {663},
          eid = {A166},
        pages = {A166},
          doi = {10.1051/0004-6361/202243607},
archivePrefix = {arXiv},
       eprint = {2206.00064},
 primaryClass = {astro-ph.SR},
       adsurl = {https://ui.adsabs.harvard.edu/abs/2022A&A...663A.166B}
}

@ARTICLE{Vogler2007,
       author = {{V{\"o}gler}, A. and {Sch{\"u}ssler}, M.},
        title = "{A solar surface dynamo}",
      journal = {\aap},
         year = 2007,
        month = apr,
       volume = {465},
       number = {3},
        pages = {L43-L46},
          doi = {10.1051/0004-6361:20077253},
archivePrefix = {arXiv},
       eprint = {astro-ph/0702681},
 primaryClass = {astro-ph},
       adsurl = {https://ui.adsabs.harvard.edu/abs/2007A&A...465L..43V}
}

@ARTICLE{Lites2014,
       author = {{Lites}, Bruce W. and {Centeno}, Rebecca and {McIntosh}, Scott W.},
        title = "{The solar cycle dependence of the weak internetwork flux}",
      journal = {\pasj},
         year = 2014,
        month = dec,
       volume = {66},
          eid = {S4},
        pages = {S4},
          doi = {10.1093/pasj/psu082},
       adsurl = {https://ui.adsabs.harvard.edu/abs/2014PASJ...66S...4L}
}

@ARTICLE{Rempel2014,
       author = {{Rempel}, M.},
        title = "{Numerical Simulations of Quiet Sun Magnetism: On the Contribution from a Small-scale Dynamo}",
      journal = {\apj},
         year = 2014,
        month = jul,
       volume = {789},
       number = {2},
          eid = {132},
        pages = {132},
          doi = {10.1088/0004-637X/789/2/132},
archivePrefix = {arXiv},
       eprint = {1405.6814},
 primaryClass = {astro-ph.SR},
       adsurl = {https://ui.adsabs.harvard.edu/abs/2014ApJ...789..132R}
}

@ARTICLE{Danilovic2010,
       author = {{Danilovic}, S. and {Sch{\"u}ssler}, M. and {Solanki}, S.~K.},
        title = "{Probing quiet Sun magnetism using MURaM simulations and Hinode/SP results: support for a local dynamo}",
      journal = {\aap},
         year = 2010,
        month = apr,
       volume = {513},
          eid = {A1},
        pages = {A1},
          doi = {10.1051/0004-6361/200913379},
archivePrefix = {arXiv},
       eprint = {1001.2183},
 primaryClass = {astro-ph.SR},
       adsurl = {https://ui.adsabs.harvard.edu/abs/2010A&A...513A...1D}
}

@ARTICLE{Buehler2013,
       author = {{Buehler}, D. and {Lagg}, A. and {Solanki}, S.~K.},
        title = "{Quiet Sun magnetic fields observed by Hinode: Support for a local dynamo}",
      journal = {\aap},
         year = 2013,
        month = jul,
       volume = {555},
          eid = {A33},
        pages = {A33},
          doi = {10.1051/0004-6361/201321152},
archivePrefix = {arXiv},
       eprint = {1307.0789},
 primaryClass = {astro-ph.SR},
       adsurl = {https://ui.adsabs.harvard.edu/abs/2013A&A...555A..33B}
}

@ARTICLE{Lagg2025,
       author = {{Korpi-Lagg}, Andreas and {Gandorfer}, Achim and {Solanki}, Sami K. and {del Toro Iniesta}, Jose Carlos and {Katsukawa}, Yukio and {Bernasconi}, Pietro and {Berkefeld}, Thomas and {Feller}, Alex and {Riethm{\"u}ller}, Tino L. and {{\'A}lvarez-Herrero}, Alberto and {Kubo}, Masahito and {Mart{\'\i}nez Pillet}, Valent{\'\i}n and {Smitha}, H.~N. and {Orozco Su{\'a}rez}, David and {Grauf}, Bianca and {Carpenter}, Michael and {Bell}, Alexander and {{\'A}lvarez-Alonso}, Mar{\'\i}a-Teresa and {{\'A}lvarez Garc{\'\i}a}, Daniel and {Aparicio del Moral}, Beatriz and {Ati{\'e}nzar}, Julia and {Ayoub}, Daniel and {Bail{\'e}n}, Francisco Javier and {Bail{\'o}n Mart{\'\i}nez}, Eduardo and {Balaguer Jim{\'e}nez}, Maria and {Barthol}, Peter and {Bayon Laguna}, Montserrat and {Bellot Rubio}, Luis R. and {Bergmann}, Melani and {Blanco Rodr{\'\i}guez}, Julian and {Bochmann}, Jan and {Borrero}, Juan Manuel and {Campos-Jara}, Antonio and {Castellanos Dur{\'a}n}, Juan Sebasti{\'a}n and {Cebollero}, Mar{\'\i}a and {Conde Rodr{\'\i}guez}, Aitor and {Deutsch}, Werner and {Eaton}, Harry and {Fern{\'a}ndez-Medina}, Ana Belen and {Fernandez-Rico}, German and {Ferreres}, Agustin and {Garc{\'\i}a}, Andr{\'e}s and {Garc{\'\i}a Alarcia}, Ram{\'o}n Mar{\'\i}a and {Garc{\'\i}a Parejo}, Pilar and {Garranzo-Garc{\'\i}a}, Daniel and {Gasent Blesa}, Jos{\'e} Luis and {Gerber}, Karin and {Germerott}, Dietmar and {Gilabert Palmer}, David and {Gizon}, Laurent and {G{\'o}mez S{\'a}nchez-Tirado}, Miguel Angel and {Gonz{\'a}lez-B{\'a}rcena}, David and {Gonzalo Melchor}, Alejandro and {Goodyear}, Sam and {Hara}, Hirohisa and {Harnes}, Edvarda and {Heerlein}, Klaus and {Heidecke}, Frank and {Heinrichs}, Jan and {Hern{\'a}ndez Exp{\'o}sito}, David and {Hirzberger}, Johann and {Hoelken}, Johannes and {Hyun}, Sangwon and {Iglesias}, Francisco A. and {Ishikawa}, Ryohtaroh T. and {Jeon}, Minwoo and {Kawabata}, Yusuke and {Kolleck}, Martin and {Laguna}, Hugo and {Lomas}, Julian and {L{\'o}pez Jim{\'e}nez}, Antonio C. and {Manzano}, Paula and {Matsumoto}, Takuma and {Mayo Turrado}, David and {Meierdierks}, Thimo and {Meining}, Stefan and {Monecke}, Markus and {Morales-Fern{\'a}ndez}, Jos{\'e} Miguel and {Moreno Mantas}, Antonio Jes{\'u}s and {Moreno Vacas}, Alejandro and {M{\"u}ller}, Marc Ferenc and {M{\"u}ller}, Reinhard and {Naito}, Yoshihiro and {Nakai}, Eiji and {N{\'u}{\~n}ez Peral}, Armon{\'\i}a and {Oba}, Takayoshi and {Palo}, Geoffrey and {P{\'e}rez-Grande}, Isabel and {Piqueras Carre{\~n}o}, Javier and {Preis}, Tobias and {Przybylski}, Damien and {Quintero Noda}, Carlos and {Ramanath}, Sandeep and {Ramos M{\'a}s}, Jose Luis and {Raouafi}, Nour and {Rivas-Mart{\'\i}nez}, Mar{\'\i}a-Jes{\'u}s and {Rodr{\'\i}guez Mart{\'\i}nez}, Pedro and {Rodr{\'\i}guez Valido}, Manuel and {Ruiz Cobo}, Basilio and {S{\'a}nchez Rodr{\'\i}guez}, Antonio and {Sanchez Toledo}, Mariano and {S{\'a}nchez G{\'o}mez}, Antonio and {Sanchis Kilders}, Esteban and {Sant}, Kamal and {Santamarina Guerrero}, Pablo and {Schulze}, Erich and {Shimizu}, Toshifumi and {Silva-L{\'o}pez}, Manuel and {Singh}, Kunal and {Siu-Tapia}, Azaymi L. and {Sonner}, Thomas and {Staub}, Jan and {Strecker}, Hanna and {Tobaruela}, Angel and {Torralbo}, Ignacio and {Tritschler}, Alexandra and {Tsuzuki}, Toshihiro and {Uraguchi}, Fumihiro and {Volkmer}, Reiner and {Vourlidas}, Angelos and {Vukadinovi{\'c}}, Du{\v{s}}an and {Werner}, Stephan and {Zerr}, Andreas},
        title = "{SUNRISE III: Overview of Observatory and Instruments}",
      journal = {\solphys},
         year = 2025,
        month = may,
       volume = {300},
       number = {5},
          eid = {75},
        pages = {75},
          doi = {10.1007/s11207-025-02485-1},
archivePrefix = {arXiv},
       eprint = {2502.06483},
 primaryClass = {astro-ph.IM},
       adsurl = {https://ui.adsabs.harvard.edu/abs/2025SoPh..300...75K}
}

@ARTICLE{delToro2025,
       author = {{del Toro Iniesta}, J.~C. and {Orozco Su{\'a}rez}, D. and {{\'A}lvarez-Herrero}, A. and {Sanchis Kilders}, E. and {P{\'e}rez-Grande}, I. and {Ruiz Cobo}, B. and {Bellot Rubio}, L.~R. and {Balaguer Jim{\'e}nez}, M. and {L{\'o}pez Jim{\'e}nez}, A.~C. and {{\'A}lvarez Garc{\'\i}a}, D. and {Ramos M{\'a}s}, J.~L. and {Cobos Carrascosa}, J.~P. and {Labrousse}, P. and {Moreno Mantas}, A.~J. and {Morales-Fern{\'a}ndez}, J.~M. and {Aparicio del Moral}, B. and {S{\'a}nchez G{\'o}mez}, A. and {Bail{\'o}n Mart{\'\i}nez}, E. and {Bail{\'e}n}, F.~J. and {Strecker}, H. and {Siu-Tapia}, A.~L. and {Santamarina Guerrero}, P. and {Moreno Vacas}, A. and {Ati{\'e}nzar Garc{\'\i}a}, J. and {Dorantes Monteagudo}, A.~J. and {Bustamante}, I. and {Tobaruela}, A. and {Fern{\'a}ndez-Medina}, A. and {N{\'u}{\~n}ez Peral}, A. and {Cebollero}, M. and {Garranzo-Garc{\'\i}a}, D. and {Garc{\'\i}a Parejo}, P. and {Gonzalo Melchor}, A. and {S{\'a}nchez Rodr{\'\i}guez}, A. and {Campos-Jara}, A. and {Laguna}, H. and {Silva-L{\'o}pez}, M. and {Blanco Rodr{\'\i}guez}, J. and {Gasent Blesa}, J.~L. and {Rodr{\'\i}guez Mart{\'\i}nez}, P. and {Ferreres}, A. and {Gilabert Palmer}, D. and {Torralbo}, I. and {Piqueras}, J. and {Gonz{\'a}lez-B{\'a}rcena}, D. and {Fern{\'a}ndez}, A.~J. and {Hern{\'a}ndez Exp{\'o}sito}, D. and {P{\'a}ez Ma{\~n}{\'a}}, E. and {Magdaleno Castell{\'o}}, E. and {Rodr{\'\i}guez Valido}, M. and {Korpi-Lagg}, Andreas and {Gandorfer}, Achim and {Solanki}, Sami K. and {Berkefeld}, Thomas and {Bernasconi}, Pietro and {Feller}, Alex and {Katsukawa}, Yukio and {Riethm{\"u}ller}, Tino L. and {Smitha}, H.~N. and {Kubo}, Masahito and {Mart{\'\i}nez Pillet}, Valent{\'\i}n and {Grauf}, Bianca and {Bell}, Alexander and {Carpenter}, Michael},
        title = "{TuMag: The Tunable Magnetograph for the SUNRISE III Mission}",
      journal = {\solphys},
         year = 2025,
        month = oct,
       volume = {300},
       number = {10},
          eid = {148},
        pages = {148},
          doi = {10.1007/s11207-025-02562-5},
       adsurl = {https://ui.adsabs.harvard.edu/abs/2025SoPh..300..148D}
}

@ARTICLE{Solanki1990,
       author = {{Solanki}, S.~K. and {Steiner}, O.},
        title = "{How magnetic is the solar chromosphere?}",
      journal = {\aap},
         year = 1990,
        month = aug,
       volume = {234},
       number = {1-2},
        pages = {519-529},
       adsurl = {https://ui.adsabs.harvard.edu/abs/1990A&A...234..519S}
}

@ARTICLE{Feller2025,
       author = {{Feller}, Alex and {Gandorfer}, Achim and {Grauf}, Bianca and {H{\"o}lken}, Johannes and {Iglesias}, Francisco A. and {Korpi-Lagg}, Andreas and {Riethm{\"u}ller}, Tino L. and {Staub}, Jan and {Fernandez-Rico}, German and {Castellanos Dur{\'a}n}, Juan Sebasti{\'a}n and {Solanki}, Sami K. and {Smitha}, H.~N. and {Sant}, Kamal and {Barthol}, Peter and {Bayon Laguna}, Montserrat and {Bergmann}, Melani and {Bischoff}, J{\"o}rg and {Bochmann}, Jan and {Bruns}, Stefan and {Deutsch}, Werner and {Eberhardt}, Michel and {Enge}, Rainer and {Goodyear}, Sam and {Heerlein}, Klaus and {Heinrichs}, Jan and {Hirche}, Dennis and {Meining}, Stefan and {Mende}, Roland and {Meyer}, Sabrina and {M{\"u}hlhaus}, Maria and {M{\"u}ller}, Marc Ferenc and {Monecke}, Markus and {Oberdorfer}, Dietmar and {Papagiannaki}, Ioanna and {Ramanath}, Sandeep and {Verg{\"o}hl}, Michael and {Vukadinovi{\'c}}, Du{\v{s}}an and {Werner}, Stephan and {Zerr}, Andreas and {Berkefeld}, Thomas and {Bernasconi}, Pietro and {Katsukawa}, Yukio and {del Toro Iniesta}, Jose Carlos and {Bell}, Alexander and {Carpenter}, Michael and {{\'A}lvarez Herrero}, Alberto and {Kubo}, Masahito and {Mart{\'\i}nez Pillet}, Valent{\'\i}n and {Orozco Su{\'a}rez}, David},
        title = "{The Sunrise Ultraviolet Spectropolarimeter and Imager: Instrument Description}",
      journal = {\solphys},
         year = 2025,
        month = may,
       volume = {300},
       number = {5},
          eid = {65},
        pages = {65},
          doi = {10.1007/s11207-025-02471-7},
archivePrefix = {arXiv},
       eprint = {2504.05416},
 primaryClass = {astro-ph.IM},
       adsurl = {https://ui.adsabs.harvard.edu/abs/2025SoPh..300...65F}
}

@ARTICLE{Iglesias2025,
       author = {{Iglesias}, F.~A. and {Feller}, A. and {Gandorfer}, A. and {Riethm{\"u}ller}, T.~L. and {Korpi-Lagg}, A. and {Solanki}, S.~K. and {Katsukawa}, Y. and {Kubo}, M. and {S{\'a}nchez Toledo}, M.},
        title = "{The SUNRISE Ultraviolet Spectropolarimeter and Imager: Standalone Polarimetric Calibration}",
      journal = {\solphys},
         year = 2025,
        month = may,
       volume = {300},
       number = {5},
          eid = {58},
        pages = {58},
          doi = {10.1007/s11207-025-02470-8},
       adsurl = {https://ui.adsabs.harvard.edu/abs/2025SoPh..300...58I}
}

@ARTICLE{Jafarzadeh2017b,
       author = {{Jafarzadeh}, S. and {Rutten}, R.~J. and {Solanki}, S.~K. and {Wiegelmann}, T. and {Riethm{\"u}ller}, T.~L. and {van Noort}, M. and {Szydlarski}, M. and {Blanco Rodr{\'\i}guez}, J. and {Barthol}, P. and {del Toro Iniesta}, J.~C. and {Gandorfer}, A. and {Gizon}, L. and {Hirzberger}, J. and {Kn{\"o}lker}, M. and {Mart{\'\i}nez Pillet}, V. and {Orozco Su{\'a}rez}, D. and {Schmidt}, W.},
        title = "{Slender Ca II H Fibrils Mapping Magnetic Fields in the Low Solar Chromosphere}",
      journal = {\apjs},
         year = 2017,
        month = apr,
       volume = {229},
       number = {1},
          eid = {11},
        pages = {11},
          doi = {10.3847/1538-4365/229/1/11},
archivePrefix = {arXiv},
       eprint = {1610.03104},
 primaryClass = {astro-ph.SR},
       adsurl = {https://ui.adsabs.harvard.edu/abs/2017ApJS..229...11J}
}

@ARTICLE{Riethmuller2008,
       author = {{Riethm{\"u}ller}, T.~L. and {Solanki}, S.~K. and {Zakharov}, V. and {Gandorfer}, A.},
        title = "{Brightness, distribution, and evolution of sunspot umbral dots}",
      journal = {\aap},
         year = 2008,
        month = dec,
       volume = {492},
       number = {1},
        pages = {233-243},
          doi = {10.1051/0004-6361:200810701},
archivePrefix = {arXiv},
       eprint = {0812.0477},
 primaryClass = {astro-ph},
       adsurl = {https://ui.adsabs.harvard.edu/abs/2008A&A...492..233R}
}

@ARTICLE{Jafarzadeh2025,
      author = {{Jafarzadeh}, Shahin and {Jess}, David B. and {Stangalini}, Marco and {Grant}, Samuel D.~T. and {Higham}, Jonathan E. and {Pessah}, Martin E. and {Keys}, Peter H. and {Belov}, Sergey and {Calchetti}, Daniele and {Duckenfield}, Timothy J. and {Fedun}, Viktor and {Fleck}, Bernhard and {Gafeira}, Ricardo and {Jefferies}, Stuart M. and {Khomenko}, Elena and {Morton}, Richard J. and {Norton}, Aimee A. and {Rajaguru}, S.~P. and {Schiavo}, Luiz A.~C.~A. and {Sharma}, Rahul and {Silva}, Suzana S.~A. and {Solanki}, Sami K. and {Steiner}, Oskar and {Verth}, Gary and {Vigeesh}, Gangadharan and {Yadav}, Nitin},
        title = "{Wave analysis tools}",
      journal = {Nat. Rev. Methods Primers},
        year = 2025,
        month = apr,
      volume = {5},
          eid = {21},
        pages = {21},
          doi = {10.1038/s43586-025-00392-0},
      adsurl = {https://ui.adsabs.harvard.edu/abs/2025NRvMP...5...21J}
}

@ARTICLE{Katsukawa2026,
       author = {{Katsukawa}, Y. and {del Toro Iniesta}, J.~C. and {Solanki}, S.~K. and {Kubo}, M. and {Hara}, H. and {Shimizu}, T. and {Oba}, T. and {Kawabata}, Y. and {Tsuzuki}, T. and {Uraguchi}, F. and {Shinoda}, K. and {Tamura}, T. and {Suematsu}, Y. and {Matsumoto}, T. and {Ishikawa}, R.~T. and {Naito}, Y. and {Ichimoto}, K. and {Nagata}, S. and {Anan}, T. and {Orozco Su{\'a}rez}, D. and {Sanchis Kilders}, E. and {Balaguer Jim{\'e}nez}, M. and {L{\'o}pez Jim{\'e}nez}, A.~C. and {Quintero Noda}, C. and {{\'A}lvarez Garc{\'\i}a}, D. and {Ramos M{\'a}s}, J.~L. and {Cobos Carrascosa}, J.~P. and {Labrousse}, P. and {Aparicio del Moral}, B. and {S{\'a}nchez G{\'o}mez}, A. and {Hern{\'a}ndez Exp{\'o}sito}, D. and {Bail{\'o}n Mart{\'\i}nez}, E. and {Morales Fern{\'a}ndez}, J.~M. and {Moreno Mantas}, A. and {Tobaruela}, A. and {Bustamante}, I. and {Bail{\'e}n}, F.~J. and {Blanco Rodr{\'\i}guez}, J. and {Gasent Blesa}, J.~L. and {Rodr{\'\i}guez Mart{\'\i}nez}, P. and {Ferreres}, A. and {Gilabert Palmer}, D. and {Piqueras Carre{\~n}o}, J. and {P{\'e}rez Grande}, I. and {Torralbo}, I. and {{\'A}lvarez-Herrero}, A. and {Korpi-Lagg}, A. and {Gandorfer}, A. and {Berkefeld}, T. and {Bernasconi}, P. and {Feller}, A. and {Riethm{\"u}ller}, T.~L. and {Smitha}, H.~N. and {Mart{\'\i}nez Pillet}, V. and {Grauf}, B. and {Bell}, A. and {Carpenter}, M.},
        title = "{The Sunrise Chromospheric Infrared Spectro-Polarimeter SCIP: An Instrument for SUNRISE III}",
      journal = {\solphys},
         year = 2026,
        month = jul,
       volume = {301},
       number = {7},
          eid = {99},
        pages = {99},
          doi = {10.1007/s11207-026-02696-0},
archivePrefix = {arXiv},
       eprint = {2603.17929},
 primaryClass = {astro-ph.SR},
       adsurl = {https://ui.adsabs.harvard.edu/abs/2026SoPh..301...99K}
}

\begin{appendix}
\section{One example of a Class 2 Ca~{\sc ii}~H brightening associated with a flux emergence event}
\label{appendix:A}
   
   \begin{figure*}
     \centering
     \includegraphics[width=0.8\hsize]{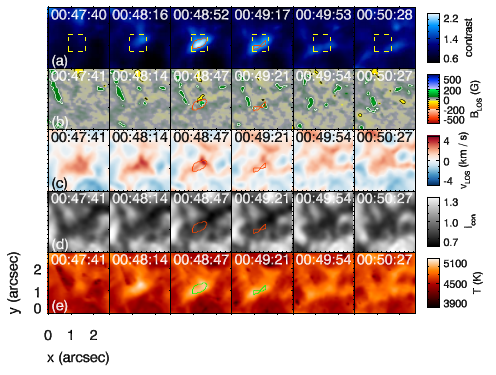}
     \caption{Same as Figure~\ref{fig:case_class2_2} but for the evolution of a Class 2 Ca~{\sc ii}~H brightening accompanied by a magnetic emergence event.}
     \label{fig:case_class2_1}
   \end{figure*}

   Figure~\ref{fig:case_class2_1} gives an example of a Ca~{\sc ii}~H brightening that is accompanied by a flux emergence event. The Ca~{\sc ii}~H intensity keeps increasing in the first three frames, and a brightening fulfilling the criteria outlined in Sect.~\ref{section:identification} is identified at 00:48:52 UT. During this period, a pair of magnetic patches with opposite polarities appears and grows, both reaching a maximum diameter of around $0.4''$ at 00:48:47 UT with a $B_{LOS}$ of about 100 G. The corresponding $v_{LOS}$ maps also show that the downflow becomes stronger during the appearance. After this period, however, the Ca~{\sc ii}~H intensity decreases, accompanied by the disappearance of the magnetic patch pair, as well as the weakening of the downflow (00:49:21 -- 00:50:27 UT). Similar to the Ca~{\sc ii}~H brightening shown in Figure~\ref{fig:case_class2_2}, no clear brightening can be seen in continuum, while the temperature at $\log\tau = -2.5$ enhances together with the appearance of the brightening.

\begin{figure}
     \centering
     \includegraphics[width=\hsize]{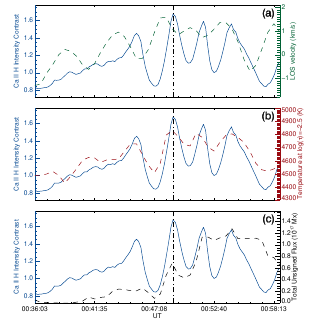}
     \caption{Same as Figure~\ref{fig:case_class22_variation} but for the Ca~{\sc ii}~H brightening case shown in Figure~\ref{fig:case_class2_1}. The vertical black dot-dashed line in each panel marks the time moment 00:48:52 UT, when the Ca~{\sc ii}~H intensity reaches the maximum.}
     \label{fig:case_class21_variation}
   \end{figure}

   The longer-term variation of Ca~{\sc ii}~H intensity, $v_{LOS}$, temperature at $\log\tau = -2.5$, and total unsigned flux in the corresponding brightening region are displayed in Figure~\ref{fig:case_class21_variation}. Similarly, the identified Ca~{\sc ii}~H brightening corresponds to just the strongest peak of the overall periodic intensity variation at 00:48:52 UT, and the $v_{LOS}$ and temperature also vary with a similar period.

   The variations in Ca~{\sc ii}~H intensity and the magnetic flux are compared in Fig~\ref{fig:case_class21_variation}c. We firstly focus on the period of time around 00:48:52 UT. Before the Ca~{\sc ii}~H intensity reaches its peak, the total unsigned magnetic flux increases by $\sim6\times10^{16}$ Mx starting from 00:47 UT, which is consistent with the appearance of the magnetic patch pair in Fig~\ref{fig:case_class2_1}b. The new flux emergence/appearance could trigger magnetic reconnection with the overlying field, which would cause localized plasma heating and in turn the excess brightening seen in the upper atmosphere. Then the magnetic flux decreases along with the drop in the Ca~{\sc ii}~H intensity. For the subsequent two periods, the magnetic flux reaches a local maximum together with the Ca~{\sc ii}~H intensity at 00:51 UT and 00:54 UT, respectively. Note that the total unsigned flux increases to around $1.1\times10^{17}$ Mx because there is a positive-polarity magnetic patch moving into the yellow box region, and cancellation happens between this patch and other tiny negative-polarity patches nearby when the Ca~{\sc ii}~H increases to its peak values.

\FloatBarrier

\section{Wavelet analysis results of the other three Ca~{\sc ii}~H contrast temporal variations}
\label{appendix:B}
   
   \begin{figure}[h!]
     \centering
     \includegraphics[width=\hsize]{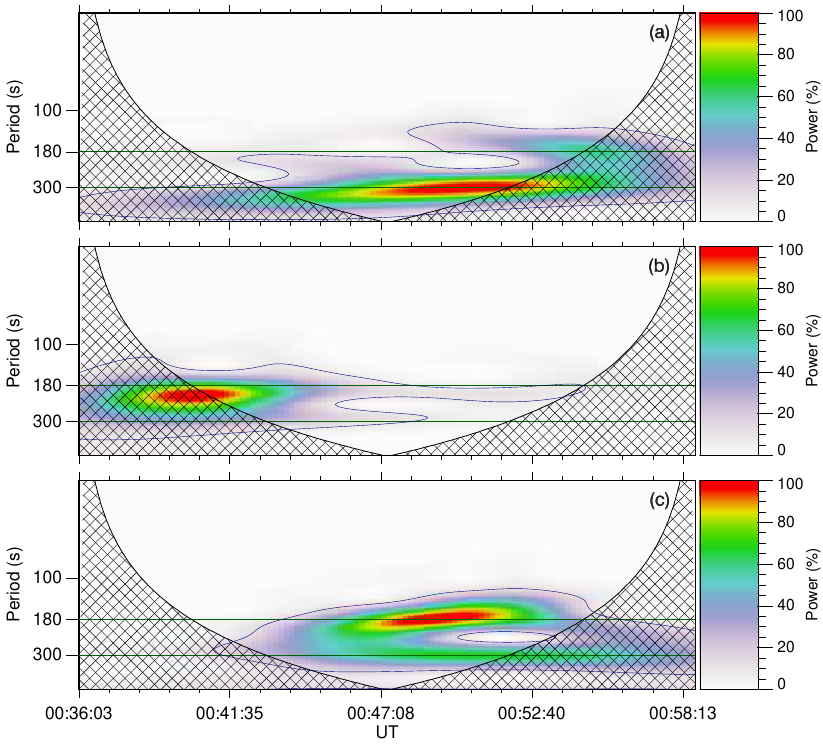}
     \caption{Wavelet power spectra of Ca~{\sc ii}~H intensity contrast variations shown in Fig~\ref{fig:case_class23_variation} ($panel \ a$), Fig~\ref{fig:case_class3_variation} ($panel \ b$), and Fig~\ref{fig:case_class21_variation} ($panel \ c$). All of the three power spectra are plotted in the same way as Fig~\ref{fig:case_class22_variation}d.}
     \label{fig:case_wavelet_gather}
   \end{figure}
   Figure~\ref{fig:case_wavelet_gather}a-c show the Morlet wavelet power spectra of Ca~{\sc ii}~H contrast variations displayed in Fig~\ref{fig:case_class23_variation}, Fig~\ref{fig:case_class3_variation}, and Fig~\ref{fig:case_class21_variation}, respectively. The signals were all linearly detrended and apodized with a Tukey window ($\alpha$ = 0.1) prior to the wave analysis. All of these three wavelet spectra indicate that the Ca~{\sc ii}~H intensity contrast oscillates with a dominant period of 3-5 min.
   
\end{appendix}

\end{document}